\documentclass[a4paper, 12pt, twoside, openright]{mythesis}

\usepackage{slashbox}
\usepackage{lmodern}

\usepackage[nottoc]{tocbibind}

\usepackage{float}
\usepackage{lscape} % Useful for wide tables or figures.
\usepackage{makecell}
\usepackage{sidecap}

\graphicspath{{figures}{example}}

\usepackage[lined,ruled,linesnumbered, vlined]{algorithm2e}
\usepackage{algorithmic}

\usepackage{booktabs} % Publication quality tables
\usepackage{multirow}
\usepackage{rotating} % sideways

\usepackage{paralist}
\usepackage{enumitem}

\usepackage{bm} % Make bold, italic math symbols
\usepackage{epsfig} % for figures
\usepackage{graphicx} % another package that works for figures
\usepackage{times}
\usepackage{mathtools}
\usepackage{textcomp, gensymb} % math symbol
\usepackage{amssymb,amsmath,amsfonts} % Short math guide for LaTeX ftp://ftp.ams.org/pub/tex/doc/amsmath/short-math-guide.pdf
\usepackage{siunitx} % SI units

\usepackage{units}
\usepackage{color}

\usepackage{comment}

\usepackage{url} % Hyphenation of URLs.
\usepackage{xcolor}
\usepackage{hyperref}
\hypersetup{colorlinks,breaklinks,
            urlcolor=[rgb]{0.918,0,0.545},
            linkcolor=[rgb]{0.710,0.180,0.141},
            citecolor=[rgb]{0,0.545,0.447}}
\usepackage{bookmark}
\usepackage{slashbox}
\usepackage{xspace}
\usepackage{microtype}

\usepackage[toc,page]{appendix}

\let\originalparagraph\paragraph
\renewcommand{\paragraph}[2][.]{\originalparagraph{#2#1}}

\newlength\paramargin
\newlength\figmargin
\newlength\secmargin

\long\def\ignorethis#1{}

\usepackage[export]{adjustbox}
\usepackage[utf8]{inputenc}
\usepackage[T1]{fontenc}
\usepackage{url}
\usepackage{booktabs}
\usepackage{amsfonts}
\usepackage{nicefrac}
\usepackage{microtype}
\usepackage{placeins}
\usepackage{multirow}
\usepackage{CJKutf8}
\usepackage{amsmath}
\usepackage{amssymb}
\usepackage{hyperref}
\usepackage{tabu}
\usepackage{longtable}
\usepackage{caption}
\usepackage{algorithmic}
\usepackage{esvect}
\usepackage{float}
\usepackage[caption=false]{subfig}
\usepackage{tcolorbox}
\usepackage{slashbox}
\usepackage{listings}

\usepackage{xcolor}

\definecolor{codegreen}{rgb}{0,0.6,0}
\definecolor{codegray}{rgb}{0.5,0.5,0.5}
\definecolor{codepurple}{rgb}{0.58,0,0.82}
\definecolor{backcolour}{rgb}{0.95,0.95,0.92}

\lstdefinestyle{mystyle}{%
    backgroundcolor=\color{backcolour},   
    commentstyle=\color{codegreen},
    keywordstyle=\color{magenta},
    numberstyle=\tiny\color{codegray},
    stringstyle=\color{codepurple},
    basicstyle=\ttfamily\footnotesize,
    breakatwhitespace=false,         
    breaklines=true,                 
    captionpos=b,                    
    keepspaces=true,                 
    numbers=left,                    
    numbersep=5pt,                  
    showspaces=false,                
    showstringspaces=false,
    showtabs=false,                  
    tabsize=2
}

\usepackage{changepage}

\usepackage{biblatex}
\usepackage{hyperref}
\hypersetup{colorlinks,breaklinks,
            urlcolor=[rgb]{0.918,0,0.545},
            linkcolor=[rgb]{0.710,0.180,0.141},
            citecolor=[rgb]{0,0.545,0.447}}

\usepackage{graphicx}
\graphicspath{{./imgs/}}

\SetKwComment{Comment}{$\triangleright$\ }{}

\begin{document}
% chktex-file 1
% chktex-file 2
% chktex-file 3
% chktex-file 9
% chktex-file 12
% chktex-file 13
% chktex-file 18
% chktex-file 44
% suppress warning 3 of chktex

\title{\textbf{Secure Aggregation Is Not All You Need: Mitigating Privacy Attacks with Noise Tolerance in Federated Learning}}
% 安全聚合非唯一所需: 論聯盟式學習法下以雜訊容忍度防御隱私攻擊

\author{ \\  \\ \\
{\it John Reuben Gilbert}\\
{\it Advisor: Professor Yen-Jen Oyang} \\ \\ \\ \\  \\ \\
{\it Graduate Institute of Computer Science and Information Engineering}\\
{\it National Taiwan University} \\
{\it Taipei, Taiwan}\\ }

% 國立臺灣大學電機資訊學院資訊工程學研究所
% 碩士論文

{\date{July 27, 2022}}

\maketitle

\frontmatter
\chapter{Abstract}\label{ch:abstract}
% chktex-file 1
% chktex-file 2
% chktex-file 3
% chktex-file 9
% chktex-file 12
% chktex-file 13
% chktex-file 17
% chktex-file 18
% chktex-file 44
% suppress warning 3 of chktex

% \begin{abstract}
  Federated learning is a collaborative method that aims to preserve data privacy while creating AI models. Current approaches to federated learning tend to rely heavily on secure aggregation protocols to preserve data privacy. However, to some degree, such protocols assume that the entity orchestrating the federated learning process (i.e., the server) is not fully malicious or dishonest. We investigate vulnerabilities to secure aggregation that could arise if the server is fully malicious and attempts to obtain access to private, potentially sensitive data. Furthermore, we provide a method to further defend against such a malicious server, and demonstrate effectiveness against known attacks that reconstruct data in a federated learning setting.
% \end{abstract}

\tableofcontents
\listoffigures
\listoftables
% \listofalgorithms

%------------------------------------
% Thesis Body -- begin
%------------------------------------

\mainmatter

\chapter{Introduction}\label{ch:introduction}
% chktex-file 1
% chktex-file 2
% chktex-file 3
% chktex-file 9
% chktex-file 12
% chktex-file 13
% chktex-file 17
% chktex-file 18
% chktex-file 44
% suppress warning 3 of chktex

% \section{Introduction}

% Data Importance
% Privacy Importance
% Federated Learning
% Vulnerabilities
% Central Limit Theorem

% \begin{tcolorbox}
% [
    % standard jigsaw,
    % title=Notation,
    % opacityback=0,  % this works only in combination with the key "standard jigsaw"
% ]
  % \begin{itemize}
      % \item $u$: users, clients
      % \item $w$: model weights
      % \item $\epsilon$: a very small number
      % \item $\delta$: a very small number
      % \item $\mathcal{A}$: an algorithm
      % \item $\mathcal{L}$: loss
      % \item $\alpha $: noise constraint
      % \item $t $: time step
      % \item $T $: total time steps
      % \item $\eta $: learning rate
  % \end{itemize}
% \end{tcolorbox}

\section{Background}

% COMPLETE: Add references to first paragraph

Deep learning, a form of artificial intelligence (AI), has provided enormous capabilities to a variety of scientific and technological fields.\cite{deeplearningbook} From object detection and classification to natural language processing, recommendation systems to autonomous vehicles and agents, computer generated images to sequence analysis, deep learning has undoubtedly changed what is possible in a variety of industries.\cite{deeplearningbook} However, in all cases where deep learning is used, the quality and availability of data is crucial to the success of deep learning models.\cite{advances-and-open-problems-FL} Regulations such as the General Data Protection Regulation (GDPR), in addition to a general desire and need to protect the confidentiality of medical, financial, geolocation, or otherwise personal data, have all resulted in a need to preserve data privacy and have limited access to data.\cite{amazon-fined, hipaa, equifax-breach, bounty-hunter-phone} Naturally, privacy-preserving methods would expand the availability of data that would be beneficial for deep learning applications.\cite{dpsgd, privacy-site, privacybook}
% Naturally, this would limit the availability of data that could be beneficial for some deep learning applications in the absence of privacy-preserving techniques.

People may be justified in not wanting to entrust their sensitive data to corporations, however. For instance, major US phone companies such as T-Mobile, Sprint, and AT\&T were found to be selling customer geolocation data to data brokers, who in turn resold location data to bail bondsmen and bounty hunters, such that anyone with a few hundred dollars USD could track the location of a phone in the US.\cite{bounty-hunter-phone} Even in cases where data is not deliberately mismanaged, it could become accessible to malicious parties in the event of a security breach, such as that of the Equifax breach in 2017, in which the personal information of 147 million people in the US were compromised.\cite{equifax-breach} Such information could be used for fraud or identity theft, which could be financially devastating to targeted individuals. The potential for data breaches can make storing large amounts of sensitive data on company servers risky. Even companies with relatively strong security and privacy policies, such as Google, are not immune to attacks, as was made evident by a breach of Google servers by Chinese hackers in 2009, which targeted Google source code as well as the email accounts of activists critical of the Chinese government.\cite{chinese-breach-google} Additionally, data access could have consequences for more than just individuals. Cambridge Analytica made use of Facebook's advertising services to amass personal information on millions of Facebook users and provide them with highly targeted political ads, thereby influencing the outcome of the 2016 US presidential election.\cite{cambridge-analytica} Violations to user privacy can also have financial consequences for companies, as Amazon was fined \texteuro746 million for violating GDPR regulations in Luxembourg.\cite{amazon-fined} Ultimately, these examples illustrate a critical need for trustworthy privacy-preserving methods for machine learning, so that both clients providing data and companies building machine learning models can trust that privacy will not compromised.

% google hacked by china
% https://www.washingtonpost.com/world/national-security/chinese-hackers-who-breached-google-gained-access-to-sensitive-data-us-officials-say/2013/05/20/51330428-be34-11e2-89c9-3be8095fe767_story.html

% amazon fined
% https://www.newsmax.com/newsfront/amazon-privacy-concern-fine/2021/08/03/id/1031045/

% that can be trusted by both clients providing data and corporations training machine learning models.
% How data is accessed by third parties can also have consequences beyond an individual level. For instance, 

% cambridge analytica election influence
% https://www.theguardian.com/news/2018/may/06/cambridge-analytica-how-turn-clicks-into-votes-christopher-wylie

% Credit reporting agencies such as Equifax and Experian that collect, aggregate, and characterize financial data on most of the US
% https://www.ftc.gov/enforcement/cases-proceedings/refunds/equifax-data-breach-settlement

% https://www.vice.com/en/article/nepxbz/i-gave-a-bounty-hunter-300-dollars-located-phone-microbilt-zumigo-tmobile

% https://www.eff.org/wp/behind-the-one-way-mirror#Data-brokers

Traditional approaches to deep learning typically involve the storage of large amounts of data on a server, on which a deep learning model can be trained directly. However, such an approach runs the risk of sensitive data becoming leaked, stolen, or mismanaged. Federated learning is a collaborative learning method originally proposed by Google with the ultimate goal of allowing AI models to be trained on private data, without any need of storing data on company servers. Rather than training a model on data stored directly on a central server, in federated learning, model weights and parameters would be sent to various clients, i.e.\@ people or organizations with their own local devices containing potentially confidential data. The local devices would then perform some degree of training on their locally stored data, in which model parameters known as weights would be gradually tuned to make the model better predict a desired output for given training input data. After training a few iterations, these updated model weights would be returned to the server to be aggregated into a global model. This process could be continued many times on many different devices, allowing for a company or organization to train a model on data to which the organization does not have direct access. The overall goal of federated learning would be to allow deep learning to occur on confidential data, without any compromises to the privacy of individuals.\cite{advances-and-open-problems-FL}

The naive approach to federated learning alone does not provide privacy guarantees, however. For example, Zhu et al.\@ have shown that mere knowledge of gradients, or the change in model weights from a training step, could be used to reproduce original training data in what is known as a reconstruction attack.\cite{deep-leakage-from-gradients} Hitaj et al.\@ have shown that in a collaborative learning setup, Generative Adversarial Networks (GANs) can be used to potentially reproduce data containing similarities to training data.\cite{deep-models-under-gan} Carlini et al.\@ have demonstrated that language models have a tendency to memorize some of the training data, and that fully trained language models can be manipulated to leak confidential information originally contained within the training data.\cite{secret-sharer} Consequently, to ensure privacy of user data, at a bare minimum the local model updates from each client device must be obscured prior to aggregation into a global model. Otherwise, the server could conceivably reconstruct original training data that was used on local client devices.

Generally speaking, approaches to obscure a model involve modifying its weights, either by adding noise or through encryption, with such approaches involving techniques such as differential privacy, secure aggregation, or homomorphic encryption.\cite{dpsgd, original-secagg, cryptonets} Differential privacy is an area of research that mathematically formalizes how strong of a privacy guarantee is obtained with the amount of noise added, with the general premise being that stronger privacy guarantees involve a tradeoff of worsened accuracy.\cite{privacybook} Secure aggregation applies secret sharing and multi-party computation protocols to a federated learning setting, such that each client can add a relatively large amount of noise to mask its local model weights, and then communicate securely with other clients so that other clients can remove a portion of that mask from their weights.\cite{original-secagg} Doing so would hide the original local models, but when aggregated, the masks would cancel so that if done correctly, the aggregated result would be the same as if the local models had been aggregated directly. Homomorphic encryption refers to encryption techniques that allow for mathematical operations to be performed on encrypted information, such that when decrypted, the result is the same as if the original unencrypted data had undergone those same mathematical operations.\cite{ckks}

A downside of sole reliance on differential privacy is that it typically involves a tradeoff with accuracy, which under some circumstances may be unacceptable, and while some progress has been made in performing deep learning on encrypted neural networks or encrypted data, current approaches of encrypted deep learning tend to face difficulties with scaling to large systems or with making use of specialized hardware such as GPUs or TPUs.\cite{privacy-site, cryptonets} Consequently in the context of federated learning, a decent amount of research has been focused on secure aggregation.

A major issue of secure aggregation, however, is that it places a degree of trust in the server to implement it correctly. Although effort was taken by researchers at Google to consider cases where there might be collusion between a small fraction of the clients and the server, they still required that the server be semi-honest and follow their protocol. They operated under the assumption that there existed a trustworthy public key infrastructure, and that the server is honest about which clients are participating in a training round. Although public key cryptography can allow messages to be passed securely to a desired recipient, it makes no guarantees as to the trustworthiness of that recipient. Consequently, it is conceivable that a malicious server could generate an arbitrarily large set of fake clients, referred to as Sybils, and perform a man-in-the-middle attack on all of the clients involved in the secure aggregation process by being dishonest about which clients are participating in a given round. This could ultimately allow the server to uncover original client weights, and conceivably then reconstruct their training data considered to be private. 

% they did not explicitly address the case where the server prompts all clients to exchange parts of masks with a set of clients owned by the server, (i.e., an attempted man-in-the-middle attack). They operated under the assumption that there existed a trustworthy public key infrastructure, and that the server is honest about which clients are participating in a training round. Although public key cryptography can allow messages to be passed securely to a desired recipient, it makes no guarantees as to the trustworthiness of that recipient. Consequently, it is conceivable that a server could generate an arbitrarily large set of fake clients, or control a subset of actual clients, and perform a man-in-the-middle attack on all of the clients involved in the secure aggregation process by being dishonest about which clients are participating in a given round. Alternatively, if the server has control over the selection process as to which clients can be selected for training, the server could intentionally favor clients that it controls to further compromise secure aggregation via its reliance on secret sharing. These methods could allow the server to obtain the original masks used by each participating client, and conceivably reconstruct the original local model weights prior to aggregation. Such information could potentially be used to then implement a reconstruction attack and obtain original client data considered to be private.

It is our belief that if a server cannot be trusted with private user data, then it's conceivable that it may not be trusted to not attempt a Sybil man-in-the-middle attack during secure aggregation. Although secure aggregation can ensure privacy when a server implements it correctly, we believe that previous work in the area of secure aggregation does not adequately address the scenario of a fully malicious server, which motivates our work.

% This brings into question the practicality of secure aggregation, and whether or not it actually provides any real security benefits as to the privacy of client data.

% An issue of solely relying on differential privacy would be that under some circumstances, an acceptable level of privacy may require an unacceptable level of accuracy, and while some re

% \section{Stochastic Gradient Descent (SGD)}

% ...

\section{Threat Model}

We assess vulnerabilities of secure aggregation under the assumption that the server orchestrating the federated learning setup may be fully dishonest and malicious, and may intentionally try to sabotage or disobey any part of the federated learning process in an attempt to obtain private user data. Our threat model assumes that the server may be more interested in obtaining user data than in training an AI model, although it may also attempt to both train a model and steal user data. In such a scenario, unsuspecting clients may participate in federated learning thinking that their private data is secure, when in reality participating may risk having their data leaked to the server if the methods used do not provide adequate security. 

% OPTIONAL: add references, maybe
We also acknowledge that some previously developed methods for securing privacy in federated learning may rely on the use of a trusted third party, such as a separate entity setting up the public key infrastructure of secure aggregation.\cite{original-secagg} We make no assumptions that any third parties may be trusted by the client anymore than the client can trust the server. Furthermore, we assume that any third party may wish to collude with the server to obtain access to private user data. We also assume that the organization running the server may have access to a reasonable amount of funds and resources, and that if the third party does not initially wish to collude, being bribed, compromised, or impersonated may still be possibilities. 

Ultimately, we investigate potential vulnerabilities in federated learning that may arise given a fully dishonest and malicious server, alongside potentially malicious third parties if they exist, and aim to protect client privacy in such a scenario.

\section{Proposed Method}

We propose the following:

% fundamentally insecure

\begin{itemize}
  \item Current approaches to secure aggregation are not adequately secure in the context of federated learning when the server is fully malicious and dishonest, (e.g., such a server may be more interested in reconstructing private data than in actually training a deep learning model, and may choose to not follow secure aggregation protocols properly).
  % \item If a trusted third party exists, several secure aggregation algorithms can be simplified to improve efficiency
  \item Noise can be added proportional to the number of clients involved in a training round, such that with an adequate amount of clients, all known reconstruction attacks fail to succeed when targeting local client models. This can be done in conjunction with other forms of differential privacy, without significantly impacting model accuracy. 
    % without any additional compromise to model accuracy.
    % \item Differential privacy can be significantly strengthened in a federated learning setting, compared to more traditional deep learning training setups, such that all known reconstruction attacks fail to succeed for image data without compromising model accuracy
    % \item Ensemble methods can effectively mitigate risks of poisoning attacks performed by clients in a federated learning setting
\end{itemize}

Given that federated learning involves aggregating many client model updates to ultimately update a global model, noise added to a local model update via differential privacy has a reduced effect on the global model's accuracy with an increase in the number of clients. This allows for a higher degree of noise to be tolerated with local clients, such that known reconstruction attacks fail to succeed when performed on any given local model. Our method of enhancing the amount of noise added in federated learning is by far easier to implement than secure aggregation, without having any risk of being circumvented via a man-in-the-middle attack, and it ultimately can prevent reconstruction attacks in federated learning. 
% Therefore, our method can be used in addition to differential privacy and secure aggregation to further enhance user privacy, even when the server is fully malicious and actively seeks to obtain private user data.

% ...

\chapter{Literature Overview}\label{ch:literature_overview}
% chktex-file 1
% chktex-file 2
% chktex-file 3
% chktex-file 9
% chktex-file 12
% chktex-file 13
% chktex-file 17
% chktex-file 18
% chktex-file 44
% suppress warning 3 of chktex

% \section{Literature Overview}

This chapter gives an overview of federated learning, a method aimed at preserving data privacy while training a neural network in a distributed setting. Privacy vulnerabilities and privacy attacks to neural networks are then discussed, followed by an overview of differential privacy, secure aggregation, homomorphic encryption, as well as other related cryptographic methods. A brief overview of man-in-the-middle attacks is given, as this pertains to one of the main vulnerabilities of secure aggregation, followed by the central limit theorem, a statistical theorem that we will use to explain our proposed alternative to secure aggregation. Related research areas are then discussed, including poisoning and inference attacks potentially carried out by malicious clients or end-users, as well as defenses to such attacks as proposed in the literature.

\section{Federated Learning}

% This section on federated learning discusses both synchronous and asynchronous optimization algorithms, as well as non-privacy related issues and current applications of federated learning.

This section gives an overview of federated learning, and discusses optimization methods, non-privacy related issues, as well as current applications of federated learning.

\subsection{Deep Learning Terminology}

Neural network models are essentially large, interconnected functions that take an input, perform linear algebra with the input and its network parameters, and produce an output. These network parameters are referred to as weights and biases, which, when combined with input values, produce outputs referred to as activations. Deep learning refers to the use of neural networks that contain multiple layers of weights and biases, with the output activations of one layer being used as the inputs to the next layer.\cite{nndl}

Consider the linear equation for a two-dimensional straight line:

\[ y = mx + b \]

For a given layer of a simple fully-connected network, a single activation, sometimes referred to as a node or neuron, can be analogous to the function of straight line, combined with another non-linear function referred to as an activation function. For a given layer $i$, the output activations $a_{i+1}$ are produced by using the outputs of the previous layer $a_{i}$ as inputs, the weights $w_i$, the bias $b_i$, and a non-linear activation function $\theta$. For an entire layer, the weights, biases, and activations would be matrices containing many numbers. Note that for the very first layer, $a_{0}$ would be the input to the neural network, and for the very last layer, $a_{i+1}$ would be the output of the network.\cite{nndl}

\[ a_{i+1} = \theta (w_i a_{i} + b_i) \]

Training a neural network refers to the iterative tuning of weights and biases so that the network produces desirable results for a given kind of input. For instance, if you want the neural network to classify images, you would want the output of the network to correspond to the type of image fed into the network as input. The process of training involves taking training data, which has known expected outputs or labels, and feeding that data into the network. The error or loss between the produced output of the network and the expected output allows for the computation of a network gradient, denoted as $\nabla f_w(x)$ below. The gradient simply indicates the direction and magnitude that the weights and biases should be tuned in order to make the network produce better results, (i.e., results closer to the expected labels for the given training input). The learning rate, another parameter denoted below as $\eta$, is used to control how much the weights and biases are changed in the direction of the gradient. Once the gradient is computed, the network parameters are tuned in the direction of the gradient, constrained by the learning rate, and then the process is repeated, with more training data fed into the network.  

\[ w \gets w - \eta \nabla f_w(x) \]

Stochastic Gradient Descent (SGD) is an optimization algorithm by which this training process occurs. As the neural network trains for many time steps, its accuracy gradually increases and the loss converges toward a minimum.\cite{nndl} 
For simplicity, we refer to tunable network parameters, i.e., weights and biases, simply as weights, denoted as $w$.

% NOTE: Aggregation techniques: FedAvg, Krum, Trimmed Mean, Adaptive FedAvg

\subsection{Federated Averaging}

Deep learning generally involves some variation of Stochastic Gradient Descent (SGD), the algorithm through which model weights are iteratively tuned to reduce the loss, or difference between the model's predicted result versus the desired true result for a given set of input data. 
% It is assumed that the reader has some familiarity with deep learning, and an in depth discussion on the variations and tunable training parameters of SGD are beyond the scope of this thesis.

% \textbf{FedAvg}
Federated Averaging (FedAvg) is the simplest form of federated learning optimization, in which SGD is performed on individual clients in parallel and then the updated weights are averaged across devices on the server, as shown in Algorithm~\ref{fedavg-algo}.\cite{advances-and-open-problems-FL} The weights are denoted as $w_{i,t}$ for client $u_i$ at time step $t$, for a total of $n$ clients, with the global model weights obtained by averaging weights of the clients:

  % \[ x_{t+1} \gets \sum_{k=1}^M \frac{1}{M} x_{t+1}^i \]
  \[ w_{t+1} \gets \sum_{i=1}^n \frac{1}{n} w_{i,t+1} \]

Each client locally updates weights with their training data performing SGD, where $\eta$ denotes the learning rate, a constant determined by the server, and $\nabla f_w(x)$ denotes the gradient computed from a loss function using back propagation.

  % \[ w \gets w - \eta\nabla f_w (x) \]

\begin{algorithm}
\DontPrintSemicolon{}
\caption{Federated Averaging (FedAvg)}\label{fedavg-algo}
\SetKwProg{server}{Server Executes:}{}{}
\SetKwProg{clientupdate}{ClientUpdate$(u_i, w_t)$:}{}{}
% \textbf{Server Executes:}\;
\server{}{initialize $w_0$\;
  \For{each round $t=1,2,\dots,T_{global}$}{$S_t \gets $ (random set of $n$ clients)\;
    \For{each client $u_i \in S$ \textbf{in parallel}}{$w_{i,t+1} \gets $ ClientUpdate$(u_i,w_t)$\;
    }
    $w_{t+1} \gets \sum_{i=1}^n \frac{1}{n} w_{i,t+1}$\;
}
}
% \Indm \textbf{ClientUpdate($i,x_)$):}\; \Indp 
  \clientupdate{}{\For{local step $j=1,\ldots,T_{local}$}{$w \gets w - \eta\nabla f_w(x)$\;
}
return $w$ to server
}
\end{algorithm}

% \textbf{FedAvgM}
% FedAvgM (momentum)
% https://arxiv.org/pdf/1909.06335.pdf

% Federated Averaging with Momentum (FedAvgM) ...\cite{fedavgm}
% \subsection{Non-IID Data}

\subsection{Asynchronous Optimizations}

% Optimization algorithms for federated learning is an active area of research, most of which is beyond the scope of this thesis. However, two are discussed below, FedAsync and FedBuff, which differ from federated averaging in that they are asynchronous.
% There are a variety of optimization algorithms based on SGD in regards to deep learning, and federated learning is no exception.
% TODO: Reference
Compared to SGD, asynchronous optimization algorithms can provide a huge efficiency benefit for federated learning given that different client devices may perform computations at different speeds.\cite{fedbuff, fedasync}
FedAsync differs from FedAvg in that clients receive the global model weights along with a time stamp. The time stamp is returned to the server along with the client update, such that the server can give faster clients with more recent weights a higher influence on the aggregated model based on a mixing parameter and staleness function, as determined by the server.\cite{fedasync}
FedBuff is another asynchronous optimization algorithm that introduces some degree of synchronicity through the use of buffers, such that training is not significantly slowed down by slower clients.\cite{fedbuff}
Synchronous algorithms such as federated averaging could potentially be slowed down from waiting for slower devices to finish their local computations. However, it is worth noting that both secure aggregation and our proposed method assume training to be synchronous, and hence performance benefits from asynchronous optimization could result in a tradeoff between efficiency and privacy.

\subsection{Technical Issues and Non-IID Data}

% 2.1 algo challenges
% 3.1 non-iid
% 3.5 communication and compression
% 6.1 data bias
% 7.1-2 system bias

% This section briefly discusses issues unrelated to privacy or security in federated learning.

% \textbf{Algorithmic Challenges} Issues such as an unreliable availability of clients,

% non-iid

\textbf{Bias and Non-IID Data} How clients are selected can introduce potential bias into training. For instance, if federated learning is performed with mobile phones, and phones must be plugged in for training to take place, this will likely introduce bias with regard to time zones and people's work and sleep schedules. Similarly, if devices are selected only at specific times, devices that are available when most other devices are not may become overrepresented in the training data. If devices are given a higher influence on the global model if they provide computed results faster, then newer devices may become overrepresented, along with wealthier regions or clients that can afford such devices or better network connectivity, which may be a particular concern when using asynchronous algorithms such as FedAsync.\cite{advances-and-open-problems-FL}

In an ideal training setting, data would be balanced, or identically and independently distributed (IID), such that changing the order in which the model views batches of data would have little or no impact on the final outcome of training. However, given the circumstances of federated learning, this cannot be guaranteed. Data can be unique to specific clients, geographic locations, or time zones, such that different kinds of data may only be available at specific times of day, leading to potentially non-IID training data. Additionally, it may be possible that different types of clients become available for training later on in the training process, and hence the distribution of data itself may change over time. Given the desire to preserve privacy, however, observing training data to ensure it is IID is not allowed, and hence it can be difficult to effectively mitigate this issue. Potential mitigation strategies could involve the use of data augmentation shared across clients, to first train with a shared public dataset prior to finetuning with federated learning, or to train multiple models.\cite{advances-and-open-problems-FL}

\textbf{Communication and Compression} The possibility of message drops and unreliable clients naturally make federated learning more technically challenging than traditional deep learning. As a result, fault tolerant algorithms, gradient compression and network quantization remain active areas of research with regard to federated learning.\cite{advances-and-open-problems-FL}
% , but are beyond the scope of this thesis.

\subsection{Current Applications}
% , most notably Google and Apple

Federated learning is currently being used in production by a variety of companies. Apple uses it in iOS 13 and above, its quicktype keyboard, and Siri. Doc.ai uses it for medical research applications. Snips uses it for hotword detection. Google uses it for its Gboard mobile keyboard, pixel phones, android phones, and android messages.\cite{advances-and-open-problems-FL} 

Additionally, Google has proposed using a variation of it in an experimental method to replace cookies in web browsers known as FLoC, or Federated Learning of Cohorts. FLoC works by the browser categorizing users based on their recent activity, and then by making available to websites only that category to which that user pertains so as to serve more personalized ads and content. However, FLoC has received heavy criticism by the Electronic Frontier Foundation (EFF), due to it being potentially much easier for websites and third parties to track and obtain personal information of users than traditional cookies.\cite{floc-terrible} Consequently, it can be argued that for federated learning to be effective at preserving privacy, the application for which it is used must also be oriented towards preserving privacy, whereas the goal of providing ads based on individual user behavior is not necessarily aligned with the goal of preserving privacy.

Applications for federated learning have been proposed in a variety of domains such as finance risk prediction, pharmaceutical drug discovery, the mining of health records, medical data segmentation, and smart manufacturing.\cite{advances-and-open-problems-FL}

% google:
% Gboard mobile keyboard [376, 222, 491, 112, 383]
% pixel phones [14]
% android messages [439]
% FLoC

% apple:
% ios 13 [25]
% quicktype keyboard [26]
% siri [26]

% doc.ai:
% FL for medical research [149]

% snips:
% hotword detection [298]

% Cross-silo applications have also been proposed or described in myriad domains including finance risk
% prediction for reinsurance [476], pharmaceuticals discovery [179], electronic health records mining [184],
% medical data segmentation [15, 139], and smart manufacturing [354]

% FLOC issues
% https://www.eff.org/deeplinks/2021/03/googles-floc-terrible-idea

\subsection{Decentralized Learning}

Decentralized learning is quite similar to federated learning, with the exception that decentralized learning involves clients sharing models in a peer-to-peer network as opposed to coordinating with a central server, and with the clients eventually converging to a global model. In such a situation, a central server may be involved in setting up the training process, such as selecting the model architecture or training hyperparameters, but otherwise the central server does not manage connections between clients.\cite{advances-and-open-problems-FL} For example, Lian et al.\@ presented AD-PSGD, an asynchronous and decentralized alternative to SGD, and demonstrated convergence.\cite{sgd-decentralized} Assran et al.\@ presented Stochastic Gradient Push (SGP) for accelerating distributed training of neural networks.\cite{sgdpush} Bellet et al.\@ provided a fully asynchronous peer-to-peer optimization algorithm for performing deep learning in a decentralized setting, and considered the addition of differential privacy for protecting client data.\cite{personalized-private-p2p} Decentralized learning may ultimately allow for better privacy guarantees than federated learning, as no single entity orchestrates the entire training process, although its asynchronous nature creates additional challenges for methods such as secure aggregation. Additionally, decentralization can result in additional technical challenges with regards to setup and model convergence.\cite{advances-and-open-problems-FL}
% Kasturi et al.\@ presented Hybrid Fusion, which involved the addition of an edge layer, an intermediate layer of server devices that communicated with clients directly, so that aggregation occured across multiple servers.\cite{hybrid-fusion}

\section{Privacy Vulnerabilities}

% The naive approach to federated learning alone does not provide privacy guarantees, however. Zhu et al. have shown that mere knowledge of gradients from a training step can be used to reproduce original training data in what is known as a reconstruction attack, and Hitaj et al. have shown that in a collaborative learning setup, Generative Adversarial Networks (GANs) can be used to potentially reproduce data containing similarities to training data.\cite{deep-leakage-from-gradients, deep-models-under-gan} Carlini et al. have demonstrated that language models have a tendency to memorize some of the training data, and that fully trained language models can be manipulated to leak confidential information originally contained within the training data.\cite{secret-sharer} Consequently, to ensure privacy of user data, at a bare minimum the local model updates from each client device must be obscured prior to aggregation into a global model. Otherwise, the server could conceivably reconstruct original training data that was used on local client devices.

This section discusses various privacy attacks that are possible on neural networks for both image-based models and language models.

\subsection{Model Inversion Attack (MIA)}

Fredrickson et al.\@ introduced an attack on machine learning systems that can reproduce features from original input training data. Experimentally they performed this attack on linear regression models, decision trees, and neural networks, although they did not assess their attack in the presence of differential privacy or other such protective methods. For facial recognition models, they were able to reproduce images that contained some resemblance to the original training data, albeit with a significant amount of noise in the reconstructed data.\cite{mia}

As shown in Algorithm~\ref{mia-algo}, their model inversion attack effectively performed gradient descent for up to $T$ iterations with stepsize $\eta$ to minimize cost produced by a cost function $C$, which involves the facial recognition model $f$ and a case-specific auxiliary cost function $\textsc{AuxTerm}$. The resulting feature vector undergoes post processing, which may involve denoising and sharpening techniques with an autoencoder neural network. The result is returned if the cost ceases to reduce after $\zeta$ iterations or if the cost reduces below the parameter $\gamma$.\cite{mia}

To our knowledge, the model inversion attack was one of the earliest demonstrations of the privacy vulnerabilities inherent in machine learning models. The general concepts of this attack were improved upon with subsequently developed attacks such as the Deep Leakage from Gradients (DLG) attack, which was able to reproduce clear images nearly identical to original training data from machine learning models. 

% For linear regression models, their model inversion attack algorithm effectively completes a target feature vectors for all possible values of $x_1$.

% C: cost function
% T: number of iterations
% eta: gradient step size
% zeta, gamma: threshold parameters
% f: classifier function

\begin{algorithm}
  \DontPrintSemicolon{}
  \caption{Model Inversion Attack for Facial Recognition}\label{mia-algo}
  \SetKwInput{Input}{Input}
  \SetKwInput{Output}{Output}
  \Input{label, $T, \zeta, \gamma, \eta$}
  $C(x) = 1 - \tilde{f}_w(x) + \textsc{AuxTerm}(x)$\;
  $x_0 \gets 0$\;
  \For{$t \gets 1, \ldots, T$}{$x_t \gets \textsc{Process}(x_{t-1} - \eta \nabla C(x_{t-1}))$\;
    \uIf{$C(x_t) \geq \text{max}(C(x_{t-1}), \ldots, C(x_{t-\zeta}))$}{\textbf{break}\;
    }
    \uIf{$C(x_t) \leq \gamma$}{\textbf{break}\;
    }
  }
  \textbf{return} $[\text{argmin}_{x_t}(C(x_t)), \text{min}_{x_t}(C(x_t))]$\;
\end{algorithm}

% https://www.cs.cmu.edu/~mfredrik/papers/fjr2015ccs.pdf
% Model Inversion Attacks that Exploit Confidence Information and Basic Countermeasures
% fredrickson et al

\subsection{Deep Leakage from Gradients (DLG)}

Zhu et al.\@ proposed a reconstruction attack known as Deep Leakage from Gradients (DLG) which was able to reveal significantly more information about the training images as compared to the original model inversion attack (MIA) of Fredrikson et al.\@ The DLG attack worked by first initializing dummy data using random Gaussian noise, and then gradually modifying it so that the gradients produced by that noise with the targeted model gradually became closer to real, known gradients of original training data, as shown in Algorithm~\ref{dlg-algo}.\cite{deep-leakage-from-gradients}

As shown in Algorithm~\ref{dlg-algo}, Gaussian noise was indicated as $\mathcal{N}(0,1)$. Gradients $\nabla f_w$ were computed with a loss function $\mathcal{L}$, after which the difference $\mathbb{D}$ was computed between the gradients produced by the generated dummy data and the known gradients produced by the private data. The dummy data $x'$ was updated based upon $\mathbb{D}$, so as to cause the gradients to converge, which would correspond to the generation of private input data.

\begin{algorithm}
\DontPrintSemicolon{}
\caption{Deep Leakage from Gradients (DLG)}\label{dlg-algo}
\SetKwInput{Input}{Input}
\SetKwInOut{Output}{Output}
% \SetKwProg{procedure}{procedure}{}{}
  \Input{$f_w(x)$: Differential model, $w$: model weights, $\nabla f_w(w)$: gradients calculated from training data}
\Output{private training data and labels $x,y$}\;
$x'_1 \gets \mathcal{N}(0,1), y'_1 \gets \mathcal{N}(0,1)$
\Comment*{Initialize dummy inputs and labels}
  \For{$t \gets 1, \ldots, T$}{$\nabla f_{w',t} \gets \partial \mathcal{L}(f_{x',t}, y'_t)/\partial w$
\Comment*{Compute dummy gradients}
  $\mathbb{D}_t \gets \Vert \nabla f_{w',t} - \nabla f_w \Vert^2$
\Comment*{Compute gradient difference}
$x_{i+1}' \gets x_t' - \eta \nabla_{x'_t} \mathbb{D}_t, y_{i+1}' \gets y_t' - \eta \nabla_{y'_t} \mathbb{D}_t$
\Comment*{Update data to match gradients}
}
return $x'_{t+1}, y'_{t+1}$
\end{algorithm}

Zhao et al.\@ further improved upon this attack by first obtaining the ground truth label of the original training data, and followed by directly modifying the sign of the dummy data's gradients to correspond with the known ground truth label. This improvement effectively sped up the convergence of DLG, although the original DLG algorithm would still converge on reconstructing training data in cases where Zhao's method would still be effective.\cite{idlg}

% ---- START ALGO ----

% \begin{algorithm}
% \DontPrintSemicolon
% \caption{Improved Deep Leakage from Gradients (iDLG)}
% \label{idlg-algo}
% \SetKwInput{Input}{Input}
% \SetKwInOut{Ensure}{Ensure}
% \Input{$F(x;W)$: Differentiable learning model, $W$: Model parameters, $\nabla W$: Gradients produced by private training datum $(x, c)$, $N$: maximum number of iterations. $\eta$: learning rate.}
% \Ensure{$(x',c')$: Dummy datum and label}\;
% $c' \gets i$ s.t. ${\nabla W_L^i}^T \cdot \nabla  W_L^i \leq 0, \forall j \neq i$
% \Comment*{Extract ground truth label}
% $x' \gets \mathcal{N}(0,1)$
% \Comment*{Initialize the dummy datum}
% \For{$i \gets 1,...,N$}{
% $\nabla W' \gets \partial l (F(x'; W), c')/\partial W$
% \Comment*{Calculate the dummy gradients}
% $L_G = \Vert \nabla W' - W\Vert^2_F$
% \Comment*{Calculate gradient difference (loss)}
% $x' \gets x' - \eta \nabla_{x'} L_G$
% \Comment*{Update the dummy datum}
% }
% return $x'$
% \end{algorithm}

% ---- END ALGO ----

\subsection{Generative Adversarial Network (GAN) Attacks}

Generative Adversarial Networks (GANs) were first introduced by Goodfellow et al., which provided a means of training neural networks to generate data that could plausibly belong to a given training dataset.\cite{original-gans} The GAN architecture consists of a discriminator, which assesses whether or not given input data is real or generated, and a generator, which tries to fool the discriminator with generated data. Both the discriminator and generator are trained together, such that a progressively powerful discriminator will encourage the convergence of a powerful generator.\cite{deeplearningbook} GANs have since become a well-known research area in the field of deep learning, with many variations and techniques developed to improve upon Goodfellow's original design.\cite{gan-overview} 
% COMPLETED: more gan citations

Hitaj et al.\@ proposed that traditional deep learning classifiers can be used as discriminators to train GANs by targeting a particular class from the classifier. Hitaj further extended this to show that in a collaborative training setting, such as federated learning, malicious clients could potentially train a generator while contributing to a global classifier model. Such a generator could then reproduce data mimicking training data seen by the global model. The generator could then be improved by simultaneously providing malicious updates to the global model to make the global classifier confuse the targeted class with another class known by the attacker, thereby requiring the model to pay attention to more detail with the targeted class in subsequent training rounds with other clients. Hitaj claimed that such an attack could not be easily defended against with differential privacy, because the amount of noise needed for an effective defense would render the global model relatively useless.\cite{deep-models-under-gan}

Algorithm~\ref{gan-attack-algo} shows the process of the GAN attack performed on the target model $f_w$ with weights $w$, given out by the honest participant $u_2$. The target model $f_w$ is used as the discriminator $\mathcal{D}$ of a GAN to train generator $\mathcal{G}$, with the goal of $\mathcal{G}$ eventually being able to leak information regarding the training data. The classifier model $f_w$ is turned into a discriminator by targeting only a single class specific to the model, represented as $y_{true}$. The adversary $u_1$ is also assumed to be a participant in the collaborative learning setup, and consequently is assumed to have its own set of legitimate data $D$. After performing a round of training on the generator $\mathcal{G}$, the adversary uses $\mathcal{G}$ to generate fake data labeled as $y_{fake}$ and merged with dataset $D$, prior to updating the classifier $f_w$ normally.

\begin{algorithm}
\DontPrintSemicolon{}
\caption{Collaborative Training Under GAN Attack}\label{gan-attack-algo}
\SetKwInput{Input}{Input}
\SetKwInOut{Output}{Output}
  \Input{model $f_w$ with weights $w$}
  \Output{Generator model $\mathcal{G}$, updated $f_w$}\;
  \For{epoch $t = 1,2,\ldots,T$}{Adversary $u_1$ obtains model $f_w$ of a target participant $u_2$\;
  Create a GAN with $f_w$ as the discriminator $\mathcal{D}$\;
  Train the GAN's Generator $\mathcal{G}$ on $\mathcal{D}$ targeting class $y_{true}$, which is specific to $u_2$'s private data\;
  Update $\mathcal{G}$ based on the answer from $\mathcal{D}$\;
  Generate $n$-samples of class $y_{true}$ with $\mathcal{G}$\;
  Assign label $y_{fake}$ (another arbitrary label) to generated samples of class $y_{true}$\;
  Merge the generated data with the local dataset $D$ of the adversary $u_1$\;
  Train $f_w$ on $D$ and share updated weights $w$ with $u_2$
}
\end{algorithm}

% \cite{deep-models-under-gan}

\subsection{Secret Sharer}

Carlini et al.\@ demonstrated that language models can be particularly vulnerable to leaking sensitive data, and provided a method for extracting data from language models via a shortest path search algorithm. The shortest path search was used to generate strings of text that have the lowest log-perplexity, which effectively is a mathematical representation of how surprised the language model would be by a given phrase, defined below for a given sequence $x$ of length $n$ for the language model $f_w$.\cite{secret-sharer}

% The log-perplexity of a sequence $x$ would be defined as follows:

\[ LP(x_1\ldots x_n) = -\text{log}_2 \Pr{(x_1\ldots x_n \vert f_w)} = \sum_{i=1}^n \left(-\text{log}_2 \Pr{(x_i \vert f_w(x_1\ldots x_{i-1}))} \right) \]

Carlini et al.\@ additionally demonstrated the susceptibility of language models to leak private data by inserting unusual phrases into training data, and then subsequently using the fully trained model to measure the log-perplexity of those phrases compared with similar phrases that had not been present in the original training data. They then used similar techniques along with a shortest path search algorithm to uncover sensitive information from a neural network trained on the Enron dataset. They also demonstrated that differential privacy can adequately mitigate the risk of memorization, with one assumption being that intermediate model weights computed during training would be hidden from the adversary.\cite{secret-sharer}

\subsection{Linkage and Membership Attacks}

While one can attempt to remove personally identifiable information from data, the richness of data enables anonymized data to be potentially linked to datasets in which the data is not anonymized via a linkage attack. In one notable example, anonymous medical records were linked to the governor of Massachussetts by comparing overlapping fields with those in public voter registration records.\cite{privacybook}

% Membership inference attacks against machine learning models
Similarly, membership attacks involve determining whether or not a particular individual or entity was present within a private dataset. Shokri et al.\@ showed how membership attacks could be performed on machine learning models.\cite{membership-attacks} Song et al.\@ provided a similar demonstration for text-based models.\cite{data-provenance}

Differential privacy can be used to provide privacy guarantees against the leakage of information, allowing for effective protection against attacks such as linkage or membership attacks.\cite{privacybook}

\section{Differential Privacy}

Differential privacy is a mathematical paradigm for assessing how much information can be leaked about an algorithm's inputs, regardless of the algorithm. The privacy of an algorithm is typically protected by adding noise to the input data or parts of the algorithm (e.g., gradients or weights of a neural network). Formally, for a randomized algorithm $\mathcal{A}$ and any set $S$ of its outputs, and two datasets $D, D'$ that differ at most by one element, the algorithm $\mathcal{A}$ is considered $(\epsilon, \delta)$-differentially private if it satisfies the following condition:

% Two main techniques that are generally considered to enhance privacy guarantees of federated learning are differential privacy and Secure Aggregation, both of which aim to obscure model weights with noise.

% Differential privacy involves adding a certain amount of noise to model weights or data, such that the probability of data having the same output with a given algorithm does not exceed a specified threshold, referred to as $\epsilon$. Formally, $\epsilon$-differential privacy is defined as follows:

% $$\epsilon = \left(\frac{\alpha}{\beta}\right)^{2}$$

\[ \text{Pr}\left[\mathcal{A}(D) \in S \right] \leq \text{e}^\epsilon \text{Pr} \left[\mathcal{A}(D') \in S\right] + \delta \]

The parameters $\epsilon$ and $\delta$ indicate the privacy guarantees of $\mathcal{A}$. For instance, if both $\epsilon$ and $\delta$ are close to $0$, then this implies that outputs are practically identical for both $D$ and $D'$, implying that it would be impossible for an attacker to determine if any given piece of data was in fact part of dataset $D$, thereby completely guaranteeing privacy of the input data. This would also require that the algorithm $\mathcal{A}$ could not memorize any given piece of input $x$, as the algorithm would have to statistically produce the same outputs for input datasets $D$ and $D' = D - \{x\}$. Conversely, higher values of $\epsilon$ and $\delta$ can be used to infer the potential of privacy leakage in a worse-case scenario.\cite{privacybook}

One downside of differential privacy is that there can be degradation in model accuracy with increased privacy guarantees, (i.e., a lower $\epsilon$ and $\delta$), because achieving better privacy guarantees requires a higher degree of noise to be added to the data or model. As a result, differential privacy typically involves a tradeoff between privacy guarantees and model accuracy.

\subsection{Noise}

Differential privacy is implemented by adding Laplacian or Gaussian noise scaled to $1 / \epsilon$, with $\epsilon$ defined in the privacy equation above. Laplacian noise is more localized than Gaussian noise and does not require $\delta$ in the differential privacy equation above, (i.e., Laplacian noise preserves $(\epsilon, 0)$-differential privacy). As a result, more Laplacian noise needs to be added to achieve the same $\epsilon$ when compared to Gaussian noise. This often makes Gaussian noise more desirable, as increased noise often negatively affects accuracy. The $\delta$ parameter can be viewed as a way of increasing tolerance for Gaussian noise at the expense of some privacy guarantees in a worse-case but unlikely scenario, with $\delta$ corresponding to the probability of such a worse-case scenario.\cite{privacy-site}

% Laplacian vs Gaussian

% \subsubsection{Quantification of Privacy Loss}
\subsection{Composition and Closure Under Post-Processing}

Differential privacy allows for composition of multiple differentially private mechanisms. For instance, if two methods are $(\epsilon,\delta)$-differentially private, then combining them and publishing the results of both still satisfies $(2\epsilon, 2\delta)$-differential privacy. Differential privacy is also robust under post-processing, such that fixed transformations on the output of a $(\epsilon,\delta)$-differentially private mechanism do not affect its privacy guarantees.\cite{privacybook}

% \subsubsection{Group Privacy}
\subsection{Differentially Private SGD}

Differentially private stochastic gradient descent (DP-SGD) was first shown to be practical by Abadi et al.\@ by making use of gradient clipping followed by the addition of noise during each training step of SGD\@. They also kept track of the privacy loss $(\epsilon, \delta)$ as it accumuated across multiple training steps.\cite{dpsgd}

Algorithm~\ref{dpsgd-algo} shows the process of DP-SGD, with $x_i$ representing sampled input for a total of $h$ samples per training round. The parameter $\eta$ represents the learning rate and $\mathcal{L}$ represents the loss function for computing the gradient $\nabla f_w$ of the neural network $f_w$ with weights $w$. The noise scale $\xi$ is used to constrain the noise generated from the Gaussian distribution $\mathcal{N}$, as the effect of noise compounds across training steps $t \in [T]$. The parameter $\gamma$ is used as a threshold to clip or constrain the gradients, which also helps to preserve privacy.

\begin{algorithm}
\DontPrintSemicolon{}
\caption{Differentially Private SGD}\label{dpsgd-algo}
\SetKwInput{Input}{Input}
\SetKwInOut{Output}{Output}
  \Input{Data samples $\{x_1,\ldots, x_n \}$, loss function $\mathcal{L}(w) = \frac{1}{n}\sum_i \mathcal{L}(w, x_i) $. Parameters: learning rate $\eta$, noise scale $\xi$, group size $h$, gradient norm bound or clipping threshold $\gamma$. }
  \Output{$w_T$ and compute the overall privacy cost $(\epsilon,\delta)$ using a privacy accounting method}
\For{$t \in [T]$}{Take a random sample $h_t$ with sampling probability $h/n$\;
    For each $i\in h_t$, compute $\nabla f_w(x_i) \gets \nabla_{w} \mathcal{L}(w_t, x_i) $
    \Comment*{Compute Gradient}
    $\nabla \bar{f_w}(x_i) \gets \nabla f_w(x_i) / \max \left(1, \frac{\Vert \nabla f_w(x_i) \Vert_2}{\gamma}\right) $
    \Comment*{Clip Gradient}
    $\nabla \tilde{f_w}(x_i) \gets \frac{1}{h}\left(\nabla \bar{f_w}(x_i) + \mathcal{N}(0, \xi^2 \gamma^2 I) \right)$
    \Comment*{Add Noise}
    $w_{t+1} \gets w_t - \eta_t \nabla \tilde{f_w}(x_i)$
    \Comment*{Descent}}
\end{algorithm}

McMann et al.\@ built differential privacy into training a LSTM language model in federated learning, and they observed that differential privacy came at the cost of increased computation rather than decreased utility with their results. They also assessed if stronger privacy guarantees affected model bias towards predicting more commonly used words, and they found that such a bias did not occur until privacy guarantees were strong enough such that model accuracy was commpromised.\cite{diff-private-lang-models}
% TODO: Discuss this in discussion

% (from privacybook):

% protection against linkage attack and arbitrary leakage (p 10)
% quantification of privacy loss
% composition (p 26)
% closure under post-processing (p 19)
% group privacy

% adding Laplace noise (p 32)

\section{Cryptographic Methods}

Secure aggregation methods in federated learning make use of public key cryptography and secret sharing, such as Shamir's secret sharing algorithm, to secure the transfer of information between clients. Secure aggregation techniques are also heavily based on previous research regarding multi-party computation, and some methods also make use of homomorphic encryption. This section gives an overview of cryptographic methods related to secure aggregation and federated learning.

\subsection{Public Key Cryptography}

Public key cryptography, or asymmetric key cryptography, refers to any cryptographic approach where an individual user has a public key $pk$ known by everyone, and a secret key $sk$ known only by that user, such that any message $m$ can be encrypted or hidden using public key $pk$ such that it can only be decrypted or revealed by using secret key $sk$. As such, anyone can encrypt a message using $pk$ that only the user containing $sk$ can decrypt or reveal. This allows messages to be securely sent to a particular recipient, such that no one else can determine the contents of the message.

% $$m = dec_s ( enc_p(m) )$$.
\[ c = enc(m, pk) \]
\[ m = dec ( c, sk) \]

RSA is a widely used example, and it involves modular exponentiation for very large integers in which the following property holds:

\[ m \equiv (m^e)^d \text{ mod } n \]

Any message $m$ can be encrypted with the public key $(e,n)$ by computing $m^e \text{ mod } n$. A private key consists of $(d,n)$ and any encrypted ciphertext $c$ can be decrypted by computing $c^d \text{ mod } n$. The modulus $n$ is computed from two primes, e.g. $n = pq$ for primes $p$ and $q$, and $e\cdot d \equiv 1 \text{ mod } \lambda(n)$, where $\lambda(n) = LCM(p-1,q-1)$.\cite{rsa}

RSA is used in the Diffie-Hellman key exchange to securely establish a shared key between two entities, which can then be used for subsequent communication with a more efficient symmetric key cryptographic technique, such as AES, in which the two entities use the same key for encryption and decryption.\cite{rsa, diffie-hellman, aes}
% RSA is a widely used example, particularly when used in a Diffie-Hellman key exchange to securely establish a shared key between two entities, which can then be used for subsequent communication with a more efficient symmetric key cryptographic technique, such as AES, in which the two entities use the same key for encryption and decryption.

% digital signature

% RSA
% AES

% Diffie-Hellman

\subsection{Secret Sharing}

Secret sharing is a cryptographic method that allows a group of users to collectively hide information from each other, such that the information can only be revealed if a certain threshold of participants collaborate to reveal that information. With secret sharing, a secret piece of data $m$ can be divided into into $n$ segments such that it can be completely reconstructed from $k$ pieces, where $k < n$, but that complete knowledge of $k-1$ pieces reveals no information about $m$. Shamir's secret sharing is a well-known example of such a $(k,n)$ threshold scheme, which is based on polynomial interpolation in the two-dimensional plane.\cite{shamir}

With Shamir secret sharing, a 2D polynomial $\rho(x) = y$ of degree $k-1$ is constructed such that $\rho(x_i) = y_i$ for all $i$, and the secret is placed into the equation as $a_0 = m$:

\[\rho(x) = a_0 + a_1 x + \cdots + a_{k-1}x^{k-1} \]

The secret $m$ can be split into $n$ shares by evaluating the polynomial for values $1$ through $n$:

\[ m_1 = \rho(1), \ldots, m_i = \rho(i), \ldots, m_n = \rho(n) \]

Given any subset of $k$ of these values along with their corresponding indices, $m = \rho(0)$ can be solved via polynomial interpolation, but knowledge of $k-1$ or less of these values is not sufficient to solve for $m$, thereby protecting the secret if the threshold $k$ is not reached.\cite{shamir}

% \subsection{Public Key Cryptography}

% ...

% \subsection{Secure Multi-Party Computation}
%
% \ldots

\subsection{Homomorphic Encryption}

% homomorphisms + complexity theory

A homomorphism is a type-preserving map between two algebraic structures, (e.g., addition is a homomorphism on the real numbers, because adding two real numbers together produces another real number). Similarly, homomorphic encryption is a form of encryption that allows for homomorphisms on encrypted data.\cite{ckks} An encryption scheme that is homomorphic with addition and subtraction, for instance, allows encrypted data to be added or subtracted, and yields the same result regardless of whether it is done before or after decryption. For instance, given two unencrypted messages $m_1$ and $m_2$, an asymmetric encryption scheme that is homomorphic with respect to addition, and all necessary keys:

\[c_1 = enc(m_1, pk_1) \]
\[c_2 = enc(m_2, pk_2) \]
\[c_3 = c_1 + c_2 \]
\[m_3 = dec(c_3, sk_3)\]

yields the same result as:

% \[c_1 = enc(m_1, pk_1) \]
% \[c_2 = enc(m_2, pk_2) \]
% \[m_1 = dec(c_1, sk_1) \]
% \[m_2 = dec(c_2, sk_2) \]
\[ m_3 = m_1 + m_2 \]

A homomorphic encryption scheme that allows for addition, subtraction, multiplication, and division is considered to be fully homomorphic. Note that operations such as addition, subtraction, multiplication, and division may be defined differently for ciphertext than for their unencrypted counterparts, depending on the type of encryption mechanism used.\cite{ckks}

Recall that encryption and decryption for RSA was computed as follows:

\[ enc(m, e) = m^e \text{ mod } n \equiv c \]

\[ dec(c, d) = c^d \text{ mod } n \equiv m \]

Notice that for two messages $m_1$ and $m_2$ and their corresponding ciphertexts $c_1$ and $c_2$, the following property holds:

\[ enc(m_1, e)\cdot enc(m_2, e) = (m_1^e m_2^e) \text{ mod } n \equiv (m_1 m_2)^e \text{ mod } n = enc(m_1 m_2, e) \]

\[ dec(c_1, d)\cdot dec(c_2, d) = (c_1^d c_2^d) \text{ mod } n \equiv (c_1 c_2)^d \text{ mod } n = dec(c_1 c_2, d) \]

Thus, RSA is partially homomorphic with respect to multiplication, as it allows two ciphertexts $c_1$ and $c_2$ to be multiplied together to produce the encrypted form of the multiplied inputs $m_1$ and $m_2$. RSA sparked research into homomorphic encryption after being shown to be homomorphic with respect to multiplication.\cite{RSA-homomorphism} 

Paillier et al.\@ presented an additive homomorphic encryption scheme that worked with integers.\cite{paillier} Gentry et al.\@ was the first to propose a fully homomorphic encryption scheme.\cite{gentry} Fan et al.\@ presented FV, a fully homomorphic encryption scheme for integers.\cite{fv-homomorphic} Cheon et al.\@ presented HEAAN, sometimes also referred to as CKKS, as an approximate symmetric fully homomorphic encryption scheme that builds upon Learning with Errors (LWE) and works with both real and complex numbers.\cite{ckks} Dowlin et al.\@ showed that homomorphic encryption could be applied to simple neural networks with \emph{CryptoNets}, and Chou et al.\@ improved upon this scheme with \emph{Faster CryptoNets}.\cite{cryptonets, faster-cryptonets} Zhang et al.\@ applied homomorphic encryption to cross-silo federated learning with BatchCrypt, done under the assumption that clients and the server follow the proposed protocol with the intention of hiding even the global model from the server during training.\cite{batchcrypt} However, such asymmetric encryption schemes often rely on trapdoor functions, which can be computationally expensive, particularly for the training of deep learning models, and batch encryption is incompatible with sparsification techniques.\cite{trapdoor, flashe} Chen et al.\@ provided a multi-key encryption scheme for inference with cloud providers, such that both the data used at inference is hidden from the model provider, and the model is hidden from the data provider.\cite{multikey-he} 

Homomorphic encryption can be a powerful technique for keeping information private while carrying out computations. However, the security of homomorphic encryption relies on secret keys being accessible only to the owners of the encrypted information. As a result, if applied to federated learning, homomorphic encryption only preserves security during training if the server and clients are semi-honest and do not disobey intended protocols. Alternatively, homomorphic encryption is best suited for scenarios where an entity encrypts and decrypts data, but does not carry out the computation, or vice versa, and thus it could work well for inference on private data with a fully trained model. Such would be the case with encrypted machine learning as a service, with an example being the work by Chen et al.\cite{multikey-he}

% \ldots

% Downlin et al.\@ proposed \emph{CryptoNets} as a means to preserve privacy, in which neural networks could be trained directly on encrypted data.\cite{cryptonets}

% Chou et al.\@ improved upon this with \emph{Faster CryptoNets}.\cite{faster-cryptonets}

\section{Secure Aggregation}

% Secure Aggregation is a method originally proposed by researchers at Google that aims to make use of secret sharing algorithms and public key cryptography to allow clients to secretly exchange information regarding how they obscure their model weights. Doing so allows each client to generate a large degree of noise to add to model weights to mask their updates, much more so than would be allowable with differential privacy, and then, via cryptographically secure channels, give pieces of that mask to other clients for them to subtract from their model weights. Consequently, when client updates are averaged together by the server, the masks would cancel out, leaving the actual aggregated result from the original client models. Additional precautions were also proposed to deal with the possibility of some clients dropping connection. Secure Aggregation differs from differential privacy in that the final aggregated model would not contain any of the noise used in the masks, and consequently an increased degree of noise would not result in a lower model accuracy.\cite{original-secagg} , aiming to make use of secret sharing algorithms and public key cryptography to allow clients to secretly exchange information regarding how they obscure their model weights. 

% \subsection{Original SecAgg}

Secure Aggregation is a method originally proposed by researchers at Google that applies multi-party computation (MPC) to federated learning. It makes use of secret sharing and public key cryptography to allow clients to mask their local model weights from the server, such that the masks cancel when the server aggregates the updates of many clients into a global model.\cite{original-secagg}

The originally proposed secure aggregation (SecAgg) mechanism contained five rounds of communication: key advertising, key sharing, masked input collection, consistency checks, and unmasking.\cite{original-secagg}

\textbf{Key Advertising} Each participating client $u_i$ generates a Diffie-Hellman keypair $(sk_1, pk_1)$ and an individual keypair $(sk_2, pk_2)$ along with a signature of their keypairs $sg_i$, and sends the public keys and signature $(pk_1, pk_2, sg_i)$ to the server. The public keypairs are generated from secret keys as follows, using a common generator $g$ and known modulus $p$:

\[ pk_1 = g^{sk_1} \text{ mod } p \]
\[ pk_2 = g^{sk_2} \text{ mod } p \]

\textbf{Key Sharing} The server then broadcasts the public keys to all clients. The clients can then validate the correctness of the signatures. A client $u_i$ can then create a shared secret $s_{ij}$ with each other participating client $u_j$ by raising received public keys $pk_j$ to the power of $u_i$'s secret key $sk_i$. All other participating clients $u_j$ would do the same with the public key from $u_i$. Note that for simplicity, mod $p$ is omitted from the equations below.
 % for all $j \in [N]\\{i}$
% TODO: how to validate signature

\[ pk_{j} = g^{sk_{j}} \longrightarrow s_{ij} = pk_j^{sk_i} = g^{sk_j \cdot sk_i}  \]
\[ pk_{i} = g^{sk_{i}} \longrightarrow s_{ij} = pk_{i}^{sk_j} = g^{sk_j \cdot sk_i}\]

Each pair of clients $u_i, u_j$ generate pairwise masks $(M_{ij}, M_{ji})$ that cancel when added using their first set of keys (i.e., their Diffie-Hellman keypairs), where $M_{ji} = -M_{ij}$. The masks can be generated locally assuming the clients all share a common pseudorandom number generator, using the shared secret $s_{ij}$ as the seed for the generator. Each client then computes secret shares for $M_{ij}$ and $sk_2$ for the purpose of Shamir secret sharing, so that later in the event of a dropped connection, their contribution of masks can be removed from the aggregated result. Each client encrypts its shares of $M_{ij}$ and $sk_2$ using the public keys of other clients, and then forwards the shares to the server to distribute to each corresponding client. Each client then generates a second mask, $M_2$, using its second secret key $sk_2$.

% and compute a shared secret key aggreement $m_{u,i}$ for other clients $i \in U$. Each client $u$ then computes secret shares for $M_u$ and $sk_2$, encrypts them, and then forward them to the server.

\textbf{Masked Input Collection} The server sends encrypted shares that it has received to all participating clients. Each client $u_i$ then computes masked weights $c_i$ by combining its local model weights $w_i$ with its individual mask $M_2$, along with all received pairwise masks $M_{ij}$ for remaining clients $u_j$. Each client $u_i$ sends their computed masked weights $c_i$ to the server.

% shares of other clients' masks. Each client $u$ sends the computed $c_u$ to the server.

\textbf{Consistency Check} The server sends a list of clients still participating, which each client $u_i$ signs with their signature $sg_i'$ and returns it to the server. Clients abort if less than $k$ users are participating, where $k$ is the threshold for Shamir secret sharing. This step ensures that at least $k$ users are still participating.

\textbf{Unmasking} The server sends a list of remaining participants that have not dropped out along with their signatures. Clients validate the signatures. For clients that dropped connection, remaining clients send the server corresponding shares of dropped clients' pairwise masks, and remaining clients remove the second individual mask. Afterward, the server is able to aggregate all remaining masked weights, and the masks should cancel during aggregation.

Assuming that the server is honest and consistent in declaring which clients dropped connection, then the server is only able to have either the individual mask $M_2$ from remaining clients removed, with the help of those clients, or have the pairwise masks $M_{ij}$ removed for clients that dropped connection via secret sharing. Assuming an honest setup, the individual $M_2$ mask will never get removed for clients that have had their $M_{ij}$ mask removed via secret sharing. Using two sets of masks fixes an issue in which even an honest-but-curious server may incorrectly assume a relatively slow client has dropped connection, and then request secret shares from other clients prior to receiving the delayed masked weights from the slow client. If there were no individual mask, this would potentially compromise the privacy of that client, as the server could remove the single mask for that client via secret sharing and gain access to its masked weights. Using two sets of masks results in clients either giving the pairwise mask for dropped clients, or individual masks for remaining clients. This fixes that potential vulnerability, though it assumes that the server is honest and consistent as to which clients dropped connection, and it assumes that the server does not have control over a majority of clients.\cite{original-secagg}

There have since been many proposed variations of secure aggregation in attempts to improve computational efficiency, to reduce communication overhead, and to add additional features. For instance, Zhang et al.\@ proposed SAFELearning, a secure aggregation scheme that simultaneously aimed to detect poisoning attacks.\cite{safelearning}
Fereidooni et al.\@ proposed SAFELearn, a more efficient secure aggregation scheme that was compatible with both MPC and fully homomorphic encryption.\cite{safelearn}
So et al.\@ proposed Turbo-Aggregate, which put clients into groups to reduce communication overhead from $O(n^2)$ to $O(n \log n)$.\cite{turbo-agg} So et al.\@ later combined a buffered asynchronous scheme similar to FedBuff with secure aggregation that involved masks canceling out within a buffer, regardless of the training round.\cite{secure-buffered}
Schlegel et al.\@ presented CodedSecAgg, which incorporated resiliency against straggler devices into secure aggregation, so that slower devices would not significantly slow down training.\cite{codepadded} Beguier et al.\@ combined neural network compression techniques with secret sharing to reduce communication overhead.\cite{efficient-sparse-secagg} 
Bell et al.\@ provided a secure aggregation protocol with logarithmic overhead and also made use of a shuffling mechanism to help anonymize clients.\cite{single-server-secagg}
Kadhe et al.\@ presented FastSecAgg, secure aggregation protocol with $O(\ell\log n)$ computation complexity, with $\ell$ being the length of model updates and $n$ being the number of clients. FastSecAgg included a multi-secret sharing scheme based on the Fast Fourier Transform, and had clients summing shares for the server to reconstruct.\cite{fastsecagg}
Yang et al.\@ proposed LightSecAgg, which had $O(\ell \log n)$ computation complexity like FastSecAgg, but also had linear communication overhead per user, thereby increasing its efficiency.\cite{lightsecagg}
Stevens et al.\@ presented a secure aggregation mechanism based on Learning with Errors, which they claimed was robust to a malicious server, although they did not investigate how the server might break protocol. Furthermore, their protocol relied on the server declaring which clients were participating in a given round, thereby giving the server control over client interactions.\cite{efficient-dp-secagg-lwe} 
% Furthermore, the security of their protocol stemmed from their use of differentially private SGD rather than secure aggregation, as their protocol relied on the server declaring which clients were participating in a given round, thereby giving the server control over client interactions.\cite{efficient-dp-secagg-lwe}

All of these methods assumed that the majority of clients were not fake, that the server did not maliciously break protocol, or that there was a reliable public key infrastructure (i.e., a trusted third party) that somehow prevented the server from using fake clients to compromise client privacy.

So et al.\@ showed how secure aggregation could leak private information over the course of multiple training rounds, in part due to partial user selection and participation, and partly due to how the global model carries information across training rounds. They proposed a user selection strategy that took this privacy leakage into account, thereby allowing for a secure aggregation protocol that would mitigate privacy leakage between training rounds for a semi-honest server.\cite{multiround-secagg}

Pasquini et al.\@ provided two attacks in which a fully malicious server could elude secure aggregation in federated learning: gradient suppression and a canary gradient. Both attacks leverage an ability of the server to create model inconsistency, in which the server intentionally sends different models to different clients in order to leak information. With gradient suppression, the server can construct a malicious model that produces zeroed gradients upon training. By sending honest parameters to a target client and malicious parameters to the rest, the output of secure aggregation will yield the gradient of only the target client, thereby compromising that client's privacy despite secure aggregation. With a canary gradient attack, the server can modify the model such that a tiny subset of its parameters will change only if certain data is present in the client's input data. By only including those parameters in a target client's model, the server can ascertain if certain data was present in the client's training data by observing changes in those specific parameters. Pasquini et al.\@ claimed that these attacks would work on any secure aggregation protocol. Furthermore, they claimed that the assumptions of multi-party computation are not compatible with federated learning, and that secure aggregation merely provides a false sense of security.\cite{eluding-secagg}

\section{Man-in-the-Middle Attacks}

Man-in-the-middle attacks involve an attacker posing as another party in order to intercept and access the contents of a message intended for that other party, prior to forwarding the message to the intended recipient. If the attacker succeeds, both honest parties think that the attacker's public key belongs to the other honest party, and thus the attacker is able to bypass any encryption performed on messages between those two parties.\cite{cryptobook}

Man-in-the-middle attacks have been traditionally viewed in the context of web security, with a widespread solution being websites using certificates issued by certificate authorities, which are trusted third parties that provide verification that users have established a secure connection with a given website. While the presence of certificate authorities increases the difficulty of attackers performing man-in-the-middle attacks, they are not infallible. There have been cases of certificate authorities incorrectly issuing certificates, with one particularly serious example being when a certificate for Google was fraudulently issued to the Iranian government instead of Google by a compromised certificate authority, allowing Iran to spy on traffic to Google within Iran until the certificates were later revoked.\cite{cryptobook, cert-auth-breach}
 % in 2011

\section{Central Limit Theorem}

We make use of the central limit theorem to explain how more noise can be tolerated in federated learning, compared to a more traditional setting in which the data would be owned by a single entity.

The central limit theorem maintains that when independent random variables are summed, their normalized sum tends toward a normal distribution as the sample size increases. The central limit theorem maintains that, for $n$ independent random samples $x_1, x_2, \ldots, x_n$ drawn from a distribution with overall mean $\mu$ and variance $\sigma^2$, then the distribution of the samples converges to the standard normal $Z$ as shown, where $\bar{x_n}$ represents the sample mean.\cite{central-limit}

\[ Z = \text{lim}_{n \rightarrow \infty} \sqrt{n}\left(\dfrac{\bar{x}_n - \mu}{\sigma} \right) \]

% \section{Law of Large Numbers}

% ...

% \section{Proposed Method}
%
% We propose the following:
%
% \begin{itemize}
    % \item Secure Aggregation is fundamentally insecure in the context of federated learning
    % \item Differential Privacy can be significantly strengthened in a federated learning setting, compared to more traditional deep learning training setups, such that all known reconstruction attacks fail to succeed for image data without compromising model accuracy
    % % \item Ensemble methods can effectively mitigate risks of poisoning attacks performed by clients in a federated learning setting
% \end{itemize}
%
% Given that federated learning involves averaging many client model updates to ultimately update a global model, noise added to a local model update via differential privacy has a reduced effect on the global model's accuracy with an increase in the number of clients. This allows for a higher degree of noise to be tolerated with local clients, such that known reconstruction attacks fail to succeed. Our method of enhancing the degree of differential privacy in federated learning is by far easier to implement than Secure Aggregation, without having any risk of being circumvented via a man-in-the-middle attack, and it ultimately can prevent reconstruction attacks in federated learning.

% Background
% Issues
% Proposed Method

% \section{Literature Overview}
\section{Related Research Areas}

This section provides an overview of some research into the security of AI systems as well as some optimizations of federated learning, without necessarily pertaining directly to data privacy or a malicious server. 

\subsection{Poisoning Attacks}
% analyzing federated learning through an adversarial lens
Poisoning attacks involve a malicious client, rather than a malicious server, attempting to intentionally alter the outcome of training. This can be done either by manipulating model weights or by manipulating training data. While a malicious client may simply wish to ruin the accuracy of the model, this can easily be detected and corrected by the server. A far more dangerous attack involves the use of a backdoor, in which the global model appears to have reasonable accuracy under normal circumstances, but misbehaves under specific conditions known only to the attacker. Bhagoji et al.\@ demonstrated how to effectively perform model poisoning attacks in federated learning, and furthermore how to obscure those attacks such that they are statistically no different from benign updates from regular clients.\cite{analyzing-through-adv-lens} They showed that by predicting how regular updates look from non-malicious clients, an attacker can perform \textit{boosting} to counteract and remove the effects of other client updates, and by using an alternating minimization strategy in which both training loss and the adversarial objective are alternately optimized, an adversary can make its attacks statistically undetectable. Such work highlights the need for improved defense strategies against poisoning attacks, as well as the effectiveness of poisoning attacks under constrained conditions.\cite{analyzing-through-adv-lens} 

% weight poisoning attacks on pretrained models
Kurita et al.\@ assessed the potential for poisoning attacks on pretrained models, and concluded that poisoned pretrained models could maintain the effects of poisoning after fine-tuning, that such models could be hard to differentiate from non-poisoned models, and that the reliability of pretrained models should be seriously considered prior to use.\cite{weight-poisoning-attacks-pretrained-models}

\subsection{Byzantine-Tolerant Aggregation Techniques}

Averaged values can be heavily influenced by outliers, and consequently the use of averaging for aggregation in federated learning is not particularly robust to poisoning attacks or bias. Research into aggregation techniques that are fault tolerant, or resilient to Byzantine failures, can potentially increase the difficulty of a poisoning attack and additionally add more stability to the training process. The general premise of a Byzantine-resilient algorithm is that it remains unaffected by a certain degree of failure among clients, with the failing clients referred to as Byzantine workers or machines, regardless of the cause of failure (e.g., due to a software bug, network issue, or a malicious attack).

Byzantine-tolerant aggregation methods provide a fault tolerant alternative to averaging for aggregation. Krum and Coordinate-wise Trimmed Mean work by filtering outliers.\cite{krum, trimmed-mean-median} Bulyan recursively ensures each coordinate is agreed upon by a majority of vectors prior to applying another aggregation rule.\cite{bulyan} Robust Federated Averaging (RFA) employs the use of a median to approximate the mean.\cite{robust-fedavg} Adaptive FedAvg makes the use of a Hidden Markov model to assess the quality of model updates, although this makes it incompatible with with secure aggregation.\cite{adaptive-fedavg} Karimireddy et al.\@ also showed that median-based techniques were vulnerable to other kinds of attacks. They proposed the Centered Clip method, in which they advocated for aggregating momentum instead of model weights, along with an aggregation rule that applies clipping to averaging.\cite{byzantine-history}

More details regarding Byzantine-tolerant aggregation techniques can be found in Appendix~\ref{byzantine_techniques}.

% , which aim to achieve this by employing the use of a median, a method of filtering outliers, or

% Methods such as Krum, Robust Federated Averaging, Bulyan, Coordinate-wise Trimmed Mean, Adaptive FedAvg, and Centered Clip.

\subsection{Backdoor Defenses}

% weight pruning defense, mitigating backdoor attacks in federated learning
Wu et al.\@ demonstrated that certain backdoor or poisoning attacks, particularly ones that involve changing a single pixel of an image, correspond to relatively large weights within a neural network. They were able to demonstrate that weight pruning could therefore be an effective defense against some forms of poisoning attacks.\cite{mitigating-backdoor-attacks-pruning}

% scalable private learning with PATE
Papernot et al.\@ proposed a relatively novel approach referred to as Private Aggregation of Teacher Ensembles (PATE) in order to train a network with sensitive data while preserving the privacy of the network. The overall premise of PATE is that a dataset containing sensitive data could be split into subsets, each of which would then be used to train a separate model. The aggregate ensemble of the models would then be used to label non-sensitive data, which would then be used to train a final model. This approach is a form of knowledge distillation in which a separate model or ensemble of models acts as a "teacher" to effectively train a "student" model that mimics the original teacher. The overall premise of PATE is that if the final student model would be attacked with any form of reconstruction attack, the attack would only produce non-sensitive data.\cite{scalable-private-pate}

% provably secure federated learning against malicious clients
Although PATE was originally intended as a defense against reconstruction attacks, Cao et al.\@ provided a theoretical assessment for using ensembles to defend against poisoning attacks in a federated learning setting, indicating that PATE's training architecture could also be useful in mitigating poisoning attacks. Cao et al.\@ used ensembles with the majority vote from the models being used to make predictions. In doing so, clients would have limited access to only a fraction of the global ensemble, implying that for a poisoning attack to succeed, the same attack would have to be implemented across many different clients to effectively poison the majority of the ensemble.\cite{provably-secure-fed}

\subsection{Inference Attacks}

% Adversarial attacks are attacks performed at inference-time on a fully trained and deployed machine learning model.
% adv examples are not bugs, they're features
In addition to attacks that can be performed during model training, many neural networks are vulnerable to adversarial attacks performed at inference time when a fully trained model is used to make predictions or classifications on real-world data. In such attacks, adversarial perturbations that are nearly imperceptible to humans are added to input data such that the model misclassifies the data with high confidence. While it is still debated why adversarial perturbations have the effect that they do, Ilyas et al.\@ provided evidence that adversarial examples are influenced by higher level features perceptible to a neural network but relatively imperceptible to the human eye.\cite{adv-bugs}
% adversarial examples are not easily detected
Carlini et al.\@ demonstrated that adversarial attacks at inference time are also very difficult to detect and defend against, and that many proposed defenses can be easily circumvented or broken if the attacker is aware of the defense. They concluded that one of the only reliable defenses currently available for inference-time adversarial attacks is the use of adversarial training, in which adversarial data is generated and then correctly classified and used to fine-tune or train a model. The downside of this defense is it typically results in a significant reduction in model accuracy, however.\cite{adv-examples-not-easily-detected}
% smooth adversarial examples
Zhang et al.\@ showed that such adversarial attacks in images can be further smoothed such that the perturbation only appears in relatively noisier parts of the image, thereby making the attack even more imperceptible.\cite{smooth-adv-examples}

% obfuscated gradients give a false sense of security
Adversarial examples designed to fool networks at inference time are typically produced by perturbing valid input samples while targeting a specific network to modify the output, using techniques such as the Fast Gradient Sign Method (FGSM), Projected Gradient Descent (PGD), or Carlini and Wagner (C\&W) attack.\cite{towards-evaluating-robustness, towards-deep-learning-models-resistant} Many defenses have been proposed that involve obfuscating network gradients to make creating adversarial examples more difficult. However, Athalye et al.\@ showed that commonly proposed defenses for adversarial examples that obfuscate or mask gradients can be easily bypassed via gradient approximation techniques for inference-based attacks. Their proposed attack technique, Backward Pass Differential Approximation (BPDA), involved computing the forward pass normally and approximating the backward pass with an approximation function. In doing so, networks that were modified to have broken, noisy, or hidden gradients were able to be approximated, such that effective adversarial examples could be produced to directly target those networks.\cite{obfuscated-gradients-false-sense-security}

Adversarial training remains one of the most reliable defenses against such attacks, and ultimately the degree to which it is used involves a trade-off between model accuracy and robustness against adversarial attacks.\cite{adv-examples-not-easily-detected} Additionally, in cases where an AI product is used to help make critical decisions, it may be necessary for someone to oversee the final decision-making process to further mitigate the effects of adversarial attacks.

\chapter{Methods}\label{ch:methods}
% chktex-file 1
% chktex-file 2
% chktex-file 3
% chktex-file 9
% chktex-file 12
% chktex-file 13
% chktex-file 17
% chktex-file 18
% chktex-file 44
% suppress warning 3 of chktex

% COMPLETE: overview paragraph <---

The following section discusses various ways that Secure Aggregation could be potentially compromised, followed by a proof demonstrating that the assumptions needed for Secure Aggregation are incompatible with a malicious server in federated learning. Finally, we provide details regarding our experimental setup to demonstrate the effectiveness of an alternative method to Secure Aggregation.

% \section{Methods}

\section{Breaking Secure Aggregation}

We provide proof-of-concept methods in which Secure Aggregation could potentially be bypassed by a malicious server below.

% [insert diagram here]

\subsection{Man-In-The-Middle Attack}

This attack assumes that the server is able to lie to individual clients as to which other clients are participating in a given round of training.

\begin{itemize}
  \item During the setup phase, the server registers fake clients or obtains control of legitimate clients, such that the server completely controls at least $n$ clients registered in the public key infrastructure, where $n$ is the number of users selected for a round of training 
  \item In the client selection phase, in which the server selects a subset of clients to undergo a round of training, the server randomly selects $n$ clients for training. The server then falsely informs each client that the other clients used in that round have the public keys of the $n$ clients under the server's control.
  \item During the Diffie-Hellman key exchange and signature verification steps, the server can then implement a man-in-the-middle attack by first decrypting messages with the $n$ known secret keys, prior to re-encrypting them and forwarding them to legitimate clients not under its control.
  \item With a successful man-in-the-middle attack, the server can then view shares of masks sent between legitimate clients. The server is then able to reconstruct original masks for each client using those shares.
  \item The server can then unmask the updates prior to aggregation, and then perform a reconstruction attack with local client weights to obtain private user data
\end{itemize}

\subsection{Compromising Secret Sharing}

This attack assumes the server is able to control which clients are selected for a given round of training. Even though this selection process is supposed to be random, a malicious server with control of the selection process could intentionally select clients that it controls, and merely pretend that the selection was random.

\begin{itemize}
  \item During the setup phase, the server registers fake clients or obtains control of legitimate clients, such that the server completely controls at least $k$ clients registered in the public key infrastructure, where $k$ is the threshold used for Shamir's secret sharing
  \item In the client selection phase, in which the server selects a subset of clients to undergo a round of training, the server can select at least $k + 1$ clients for training, where $k$ of those clients belong to the server or are otherwise under the server's control.
  \item Given that the server has direct control of $k$ clients, the server can simply reconstruct user masks via Shamir's secret sharing mechanism.
  \item The server can then unmask legitimate client updates prior to aggregation, and then perform a reconstruction attack with local client weights to obtain private user data
\end{itemize}

These attacks could be potentially mitigated by transferring the client selection process to a trusted third party. However, that merely transfers trust from the server to the third party, and assumes that collusion between the server and third party is not possible. Additionally, this also assumes that the server is completely unable to impersonate a trusted third party to clients, which in turn could require the use of some kind of certificate authority (i.e., a trusted fourth party) to verify the third party. This would further assume that the cerificate authority would not collude with the server, be impersonated by the server, or otherwise be tricked by the server into giving the server trust. We do not make such assumptions in our threat model of a fully malicious server, and under such circumstances, secure aggregation could potentially be compromised. 

In the event of there being such a third party, though, the server could still potentially compromise privacy by strategically dropping client connections to maintain control of a majority of clients each round.

\subsection{Strategically Dropping Connections}

This attack is a variation of Compromising Secret Sharing, but it assumes that the assumptions of the previous two attacks are not viable. For instance, if a trusted-third party declares to each client who is participating in a training round, then the man-in-the-middle attack may not be viable. If the trusted-third party selects which clients participate in a round, then the aforementioned method of compromising secret sharing may not work. However, this attack assumes that the server is still able to declare which clients have dropped connection with it, which is a property of all previously discussed secure aggregation methods. Note, however, that this attack has the highest chance of not succeeding for any particular training round.

\begin{itemize}
  \item During the setup phase, the server registers a significant amount of fake clients, such that the server completely controls a vast majority of clients registered in the public key infrastructure.
  \item If the trusted third party selects only clients controled by the server, the server pretends the round is completed.
  \item Otherwise, the server drops connection with enough legitimate clients so as to have control over a majority of clients in the round.
  \item If enough clients remain in the round to proceed, the server can compromise client masks via Shamir's secret sharing, as the server controls a majority of clients. If not enough clients remain in the round, the server pretends the round is completed.
  \item If the round proceeds, the server can unmask legitimate client updates prior to aggregation, and then perform a reconstruction attack with local client weights to obtain private user data
\end{itemize}

\section{The Insecurity of Secure Aggregation: A Proof}
% \subsection{The Problem with Secure Aggregation}

Multi-party computation (MPC) allows for a group of participants to securely compute a function without seeing the inputs. Most MPC protocols rely on secret sharing, which is secure assuming that a majority of participants are not malicious or collude. For instance, Shamir secret sharing is secure if there are $k < \frac{n}{2}$ passive adversaries or $k < \frac{n}{3}$ active adversaries where $n$ is the total number of participants. These MPC protocols also assume that participants are fully aware of who is participating, and are able to send messages to each other securely.\cite{mpc, shamir}

Two-party computation (2PC) is a subset of multi-party computation that allows for secure computation when there are only two participants. Such 2PC protocols cannot rely on secret sharing techniques for security, but instead, they tend to rely on converting the function into a boolean circuit, with Yao's garbled circuit protocol being one example.\cite{2pc} 2PC has an advantage over MPC in that it does not need a majority of honest non-malicious participants to be secure. However, it requires first converting the function to be computed into a boolean circuit, making it challenging to apply to aggregating the floating point weights of a neural network.

Secure aggregation is the application of MPC protocols to federated learning, so as to protect the privacy of client data. For each round of training, federated learning involves a central server aggregating partially trained models from many different clients into one global model.\cite{original-secagg} If MPC is used, and its assumptions hold, then the client inputs can be hidden from the server, while the aggregated output can be computed and viewed by the server, thereby allowing for a global model to be trained without the server being able to access or reconstruct training data stored on client devices. However, in the case of a malicious and dishonest server, the assumptions of MPC may not hold.

% However, in the case of a malicious and dishonest server, the assumptions of MPC may not hold. If there is no trusted third party assisting in the setup of the federated learning process, then it's the server's job to assist clients in connecting to other clients, and it's the server's job to tell clients which other clients are participating in a given training round. If the server is fully dishonest, then a man-in-the-middle (MITM) attack is possible, which would violate the assumption for MPC that clients are fully aware of who is participating. If a MITM attack were successful, each client would think that they are performing MPC with other clients, but instead they would effectively just be performing MPC directly with the server. Thus, in this scenario, secure aggregation would have to be secure for only two participants, namely an honest client and the server, which would mean that it would have to make use of 2PC, and it could not rely on secret sharing for security.

\textbf{Lemma}: \textit{If there is no trusted third party in federated learning, and if the server is malicious and dishonest, then secure aggregation cannot be secure.}

\textbf{Proof}: Suppose we have a secure aggregation protocol that uses MPC but not 2PC, and that it is robust to a fully malicious and dishonest server, even when there is no trusted third party to assist in the federated learning setup. Without a trusted third party, it becomes the server's job to select which clients participate in a training round, and it also becomes the server's job to assist the participating clients in communicating with other clients. If the server is dishonest, the server can have each client directly talk to fake clients controlled by the server, prior to forwarding the information to other clients. This constitutes a Sybil man-in-the-middle (MITM) attack, meaning the secure aggregation protocol would have to be secure for just two participants: a single honest client and the malicious server, as all client interaction would go through the server. Because the protocol uses MPC but not 2PC, this is a contradiction of initial assumptions.

Suppose instead we have a secure aggregation protocol that uses 2PC and is secure for two participants, even without a trusted third party. As previously mentioned, the absence of a trusted third party implies that this protocol is secure even when the two participants are an honest client and a malicious server. Let's define $\mathcal{A}(x_1, x_2) = y$ as the computation we wish to perform when aggregating, with $x_1$ being the input provided by the honest client, $x_2$ being the input provided by the malicious server, and $y$ being the output visible to the server. Because the malicious server is a participant, it knows both $x_2$ and $y$, and it knows the original algorithm $\mathcal{A}$, meaning that hiding $\mathcal{A}$ and its inputs with a 2PC protocol (e.g., a garbled circuit) does not preserve privacy. There is only one unknown variable, for which the server can solve. For instance, if $\mathcal{A}$ is FedAvg, then the server knows that:

\[ \mathcal{A}(x_1, x_2) = \dfrac{x_1 + x_2}{2} = y \]

Thus the server can compute the client's weights like so:

\[ x_1 = 2y - x_2 \]

Therefore the secure aggregation protocol cannot be secure for only two participants, which contradicts intitial assumptions.

As secure aggregation by definition must rely on either 2PC or MPC protocols, then, without a trusted third party, secure aggregation cannot be secure with a malicious server. This proves the initial claim.$\blacksquare$

%% qed

This requires that a trusted third party performs the setup process, and selects which clients participate in a given training round. However, if a trusted third party exists and can be trusted not to ever collude with the server, then that trusted third party could simply generate and send two masks to every client, as is originally used in secure aggregation, potentially cutting down on communication costs. Alternatively, the third party could partake in some of the aggregation, cutting down on some of the computation costs.

Even with a trusted-third party, however, secure aggregation methods often rely on the server to declare which clients dropped connection.\cite{original-secagg} By lying about which clients dropped out, the server could still maintain some control over client interactions. Furthermore, there is nothing other than trust guaranteeing that the third-party does not collude with the server. Overall, we do not assume there to be a trusted third party, but if one were to exist, it would not make secure aggregation reliably secure.
% if a trusted third party exists, that third party could replace in a portion of the secure aggregation process. If a trusted third party does not exist, which we do not assume in our threat model, then we cannot rely on secure aggregation to preserve privacy.
% Even with a trusted-third party, however, secure aggregation methods often rely on the server to declare which clients dropped connection.\cite{original-secagg} By lying about which clients dropped out, the server could still maintain some control over client interactions. Furthermore, there is nothing other than trust guaranteeing that the third-party does not collude with the server. Overall, if a trusted third party exists, that third party could partake in a portion of the secure aggregation process, thereby improving efficiency. If a trusted third party does not exist, which we do not assume in our threat model, then we cannot rely on secure aggregation to preserve privacy.

% \textbf{Compromise the Public Key Infrastructure}

% \textbf{Improving Efficiency with a Trusted Third Party}

% \ldots

\section{Masking the Model}
% \section{Enhancing Differential Privacy}
% \section{Central Limit Defense}

% COMPLETED: transition sentence, why secagg is not needed
% We believe that in most cases, Secure Aggregation is
We believe that if the number of participating clients is sufficient, then Secure Aggregation becomes unnecessary. We explain this phenomenon using the central limit theorem.

% As stated by the central limit theorem, with increasing sample size, the sum of independent random variables tends toward a normal distribution centered at $n \mu$, where $\mu$ represents the expected value and $n$ represents the total number of variables. Consequently, with a sufficiently large sample size, the average of random variables from an independent and identical distribution will also tend toward $n \mu$. Constraining the range within which random values are generated would further increase the tendency of the average to approach $n \mu$, thereby reducing the amount of samples needed to have the average approximate $n \mu$.

Let $X_1, \ldots, X_n$ be $n$ independent and identical random variables with an expected value $\mu = 0$ and standard deviation $\sigma$. Then, based on the central limit theorem, the distribution of $Y = \frac{X_1 + X_2 + \cdots + X_n}{n}$ approaches $\mathcal{N}\left(0, \frac{\sigma}{\sqrt{n}}\right)$.
% where $\sigma$ is the standard deviation of $X_i$ for $1 \leq i \leq n$.
Accordingly, as the sample size $n$ increases, $\Pr\left( -\epsilon \leq Y \leq \epsilon \right) \approx 1$, with $\epsilon$ approaching $0$. 

% _____
% the distribution $D$ can be accurately approximated by the standard normal distribution with expected value $0$, where:

% Let $X$ be a random variable with expected value $\mu$ and standard deviation $\sigma$. If such random variables are selected $n$ times with $n$ significantly exceeding $0$, then for $Y = X_1 + X_2 + \cdots + X_n$, the distribution $D$ can be accurately approximated by the standard normal distribution with expected value $n \mu$, where:

% \[ D = \dfrac{\dfrac{Y}{n} - \mu}{\dfrac{\sigma}{\sqrt{n}}} \]

% \[ \text{Prob}\left( -2.57 \leq D \leq 2.57 \right) \approx 0.99 \]

% Thus, it follows that:

% ____

% Accordingly,

% \[ \Pr\left(\mu - 2.57 \dfrac{\sigma}{\sqrt{n}} \leq Y \leq \mu + 2.57 \dfrac{\sigma}{\sqrt{n}} \right) \approx 0.99 \]

% Therefore, as the sample size $n$ increases, $\Pr\left( -\epsilon \leq Y \leq \epsilon \right) \approx 1$, with $\epsilon$ approaching $0$.
% This indicates that the average converges to $\mu$, which is $0$ for a normal distribution.

This can be applied to federated learning, such that if clients add random noise to their model weights, then the average of that noise will approximately be $0$ with a sufficient amount of clients. Similarly, constraining the noise to be within a smaller range around $0$ will further reduce the amount of clients needed to achieve an average of approximately $0$ for the added noise. We propose that a combination of constraining the noise and aggregating a sufficient number of clients causes reconstruction attacks to fail on client local models prior to aggregation, and that the additional noise will have minimal effect on the global model after aggregation. In other words, more noise can be tolerated in federated learning, regardless of model architecture or the type of data, without any need for secure aggregation.

Our defense operates under the assumption that the client is able to locally add enough noise to its local model, prior to sending it to the server, such that reconstruction attacks fail on the local model. Although unnecessary, if the client wishes, it could verify that its model is obscured by running a reconstruction attack on its masked local model, prior to sending the updated weights to the server.

The server would tell everyone its threshold for how many clients would be needed to perform an aggregation. The clients would then add noise locally to their models based on this threshold, and they could refuse to participate if the noise is not adequately high. The server would be unable to reconstruct individual client models with added noise, and the only way to reduce the effect of the noise would be to aggregate a sufficient number of clients. This reduces reliance on the server to protect privacy, and it also creates an incentive for the server to aggregate a larger number of clients, thereby further preserving privacy of local models.  

We do not assess the scenario of a malicious client that intentionally adds too much noise, but we presume such a scenario could easily be detected by the server due to a significant accuracy drop after that round of training.

Our method involves masking local model weights using uniform noise $\mathcal{U}$ generated within the range $[-\alpha, \alpha]$, where $0 \leq \alpha \leq 1$. We add the noise to local model weights prior to aggregation. 
Algorithm~\ref{enhanced-fedavg-algo} shows FedAvg without differential privacy, but with our noise added prior to aggregation. This is a modified version of Algorithm~\ref{fedavg-algo}.\cite{advances-and-open-problems-FL} 
Note that the weights are denoted as $w_{i,t}$ for client $i$ at time step $t$, for a total of $n$ clients, with $w_0$ denoting the weights initialized by the server, $\eta$ denoting the local learning rate and $\nabla f_w(x)$ denoting the local gradient.

% [insert algorithm here (reuse fedavg)]

\begin{algorithm}
\DontPrintSemicolon{}
\caption{Federated Averaging (FedAvg) with Mask}\label{enhanced-fedavg-algo}
\SetKwProg{server}{Server Executes:}{}{}
\SetKwProg{clientupdate}{ClientUpdate$(u_i, w_t)$:}{}{}
% \textbf{Server Executes:}\;
\server{}{initialize $w_0$\;
  \For{each round $t=1,2,\dots,T_{global}$}{$S_t \gets $ (random set of $n$ clients)\;
    \For{each client $u_i \in S$ \textbf{in parallel}}{$w_{i,t+1} \gets $ ClientUpdate$(u_i,w_t)$\;
    }
    $w_{t+1} \gets \sum_{i=1}^n \frac{1}{n} w_{i,t+1}$\;
}
}
% \Indm \textbf{ClientUpdate($i,x_)$):}\; \Indp 
  \clientupdate{}{\For{local step $j=1,\ldots,T_{local}$}{$w \gets w - \eta\nabla f_w(x)$\;
}
  $\tilde{w} \gets w + \mathcal{U}([-\alpha,\alpha])$\;
  return $\tilde{w}$ to server
}
\end{algorithm}

% \begin{algorithm}[H]
% \DontPrintSemicolon{}
% \caption{Enhanced Federated Averaging (FedAvg)}\label{enhanced-fedavg-algo}
% \SetKwProg{server}{Server Executes:}{}{}
% \SetKwProg{clientupdate}{ClientUpdate$(i, x_t)$:}{}{}
% \server{}{initialize $x_0$\;
% \For{each round $t=1,2,\dots,T$}{$S_t \gets $ (random set of $M$ clients)\;
    % \For{each client $i \in S$ \textbf{in parallel}}{$x_{t+1}^i \gets $ ClientUpdate$(i,x_t)$\;
    % }
    % $x_{t+1} \gets \sum_{k=1}^M \frac{1}{M} x_{t+1}^i$\;
% }
% }
% \clientupdate{}{\For{local step $j=1,\ldots,K$}{$x \gets x - \eta\nabla f(x;z)$ for $z \sim \mathcal{P}_i$\;
% }
  % $\tilde{x} \gets x + \mathcal{U}([-\alpha,\alpha])$\;
  % return $\tilde{x}$ to server
% }
% \end{algorithm}

Algorithm~\ref{enhanced-dpsgd-algo} shows Differentially Private SGD that could be performed on a local client, with the addition of noise added to weights prior to submitting weights to the server. This is a modified version of Algorithm~\ref{dpsgd-algo}.\cite{dpsgd}

As already explained for Algorithm~\ref{dpsgd-algo}, $x_i$ represents sampled input for a total of $h$ samples per training round, with $\eta$ being the learning rate, $\mathcal{L}$ being the loss function for computing the gradient $\nabla f_w(x)$ of the neural network $f_w$ with weights $w$. The noise scale $\xi$ constrains noise generated from the Gaussian distribution $\mathcal{N}$, as the effect of noise compounds across training steps $t \in [T]$. The parameter $\gamma$ is used as a threshold to clip or constrain the gradients, which also helps to preserve privacy. Additionally, $\mathcal{U}$ represents uniform noise taken between the range $[-\alpha, \alpha]$ and added to the model weights $w$, prior to sending the weights to the server.

We show Algorithm~\ref{enhanced-dpsgd-algo} as a reference as to how our method can be compatible with DP-SGD\@. While our method is very similar to DP-SGD, we have separated adding noise from the local training steps $t \in [T]$. The reason for this is that given that the noise in DP-SGD propogates across local training steps, if the noise is too high, the model $f_w$ will not converge. The main purpose of our method is to conceal clients' local models prior to aggregation, and as a result it can be advantageous to separate the noise from the local training step to allow for a faster and more stable local convergence during training, while allowing for more noise to be added to mask local models prior to aggregation. Note, however, that the majority of noise from our method is expected to be canceled out from the aggregation process. As a result, our method only aims to protect local models, as is the goal of secure aggregation, and it provides no privacy guarantees as to the aggregated global model. Consequently, it may be advantageous to use a combination of our method with DP-SGD, as the privacy guarantees from DP-SGD would carry over to protect the global model.

\begin{algorithm}
\DontPrintSemicolon{}
\caption{Differentially Private SGD with Mask}\label{enhanced-dpsgd-algo}
\SetKwInput{Input}{Input}
\SetKwInOut{Output}{Output}
  \Input{Examples $\{x_1,\ldots, x_n \}$, loss function $\mathcal{L}(w) = \frac{1}{n}\sum_i \mathcal{L}(w, x_i) $. Parameters: learning rate $\eta$, noise scale $\xi$, group size $h$, gradient norm bound or clipping threshold $\gamma$. }
  \Output{$w_T$ and compute the overall privacy cost $(\epsilon,\delta)$ using a privacy accounting method}
\For{$t \in [T]$}{Take a random sample $h_t$ with sampling probability $h/n$\;
    For each $i\in h_t$, compute $\nabla f_w(x_i) \gets \nabla_{w} \mathcal{L}(w_t, x_i) $
    \Comment*{Compute Gradient}
    $\nabla \bar{f_w}(x_i) \gets \nabla f_w(x_i) / \max \left(1, \frac{\Vert \nabla f_w(x_i) \Vert_2}{\gamma}\right) $
    \Comment*{Clip Gradient}
    $\nabla \tilde{f_w}(x_i) \gets \frac{1}{h}\left(\nabla \bar{f_w}(x_i) + \mathcal{N}(0, \xi^2 \gamma^2 I) \right)$
    \Comment*{Add Noise}
    $w_{t+1} \gets w_t - \eta_t \nabla \tilde{f_w}(x_i)$
    \Comment*{Descent}}
  $\tilde{w}_T \gets w_{T} + \mathcal{U}([-\alpha,\alpha])$
  \Comment*{Mask}
\end{algorithm}

\section{Experimental Setup}

We performed the following experiments to test the viability of our defense in the presence of known reconstruction attacks. All experiments were performed using Python 3.8 on a NVIDIA GeForce 2080 Ti GPU.\@ Code for obtaining the model architectures for each experiment can be found in Appendix~\ref{model_arch}.
% $U([-\alpha, \alpha])$ is a uniform distribution from $[-\alpha, \alpha]$

% We generated Gaussian noise between $0$ and $1$ and then multiplied it by to mask the local model weights, according to the following formula, such that $\alpha$ limited the noise between a given range:

\subsection{Noise Tolerance}

We generated each mask $M$ for local model weights using uniform noise $\mathcal{U}$ generated within the range $[-\alpha,\alpha]$ for varying values of $\alpha$, which we refer to as the noise constraint parameter:

\[ M = \mathcal{U}([-\alpha,\alpha]) \]

We assessed the accuracy of both local and global models for $10$, $100$, and $1000$ clients, while varying the noise constraint $\alpha$ between $0$ and $1$.

% As neural network weights are float values between $-1$ and $1$, we initially set $a,b$ equal to $-1$ and $1$ respectively. We later changed $a$ to $-0.9$, as this improved model stability.

% To obtain an optimal range of noise for a given network architecture in federated learning, we assessed the average accuracy of a local pretrained networks after adding the noise, and compared it to the accuracy of the aggregated global model. The goal was to maximize the distance between the accuracy of the global model and the average accuracy of local models, in order to simultaneously optimize for enhanced privacy with minimal compromise to global model accuracy. We then varied the number of clients to determine an optimal ratio of clients to $\alpha$, which would allow for an easy generalization of how much additional Gaussian noise could be tolerated given the number of clients present in a federated learning setup.

\subsection{DLG Attack}

We then performed DLG attacks on local models to attempt reconstruct original training data. We masked the local models with varying values of $\alpha$ to demonstrate the effect of the noise constraint $\alpha$ on the ability to perform DLG reconstruction attacks.
% This allowed us to determine the minimum amount of noise needed to prevent reconstruction attacks. This was then used to determine a recommended minimum number of clients for a federated learning setup, along with the minimum amount of Gaussian noise needed to protect against data reconstruction attacks.

% Additionally, we assessed the vulnerability of the global model to DLG reconstruction attacks to ensure that data reconstruction could not occur after aggregation. We performed our experiments using the CIFAR-100 dataset and pretrained ResNet architectures.

\subsection{Log-Perplexity}

We assessed the impact of varying the noise constraint $\alpha$ on log-perplexity. Log-perplexity was a critical component of the attack by Carlini et al.\@ on language models, in which a search algorithm, combined with a target language model's generative properties, was used to obtain phrases that existed in the training set.\cite{secret-sharer} A low log-perplexity indicates an increased ability for the language model to generate data similar to given input data. We used a pretrained GPT-2 language model and computed log-perplexity over the WikiText test dataset after masking the model, for varying values of $\alpha$. A higher log-perplexity implies a higher difficulty in obtaining leaked data from a masked language model.

\subsection{GAN Attack}

We assessed reconstruction attacks using a GAN based on the paper by Hitaj et al.\cite{deep-models-under-gan} We attempted to use a regular classifier as the discriminator of a GAN in order to assess the practicality of a malicious client or server performing a GAN reconstruction attack in a federated learning setting. Given that many clients contribute to a global model in federated learning, we also assessed the practicality of training a GAN using a discriminator that trained significantly faster than the generator. To mimic the possibility of an attacker initiating the attack after some training has elapsed, we also attempted training a GAN with a discriminator that was pretrained for several rounds with a different generator. We trained on the CIFAR-10 dataset with a DCGAN architecture for 5 epochs.\cite{dcgan} We then used the pretrained discriminator to train another generator from scratch for 5 epochs, as this would mimic the attack being initiated after the first training round, or after aggregation of multiple local models. We also performed training with the discriminator model being masked with uniform noise with $\alpha = 0.01$ for 5 epochs, to demonstrate the effect of masking local models with noise on the ability to carry out the GAN attack.

% \subsection{Attacking Our Method}
%
% To assess the robustness of our defense, we attempted to circumvent it by having the server generate fake models with weights set to 0, followed by adding noise equivalent to that of a client. Then the server would average the known model weights with a legitimate client's model weights. The malicious server would then attempt to reconstruct the client's original weights prior to adding noise by multiplying the averaged global weights by the total number of clients used in aggregation.
%
% The general concept of this attack can be simplified as follows.
%
% Suppose you have an unknown weight $w$ that gets averaged with any known value, $x$. If $x = 0$, then $w$ can easily be obtained by multiplying by the number of values used:
%
% $$\left(\dfrac{w + x}{2}\right) 2 = \left(\dfrac{w + 0}{2}\right) 2 = w$$
%
% However, if noise is added to both $w$ and $x$, a larger number of clients would be needed to reduce the noise. Hence this attack can be summarized as follows, where $w$ is an unknown weight from a client, $x$ is known noise generated by the server, and $n$ is an arbitrary large number of clients selected by the server:
%
%
% $$\left(\dfrac{w + \sum_{i=1}^{n} x_i}{n}\right) n \approx w$$
%
% We performed this attack for various amounts of noise and values of $n$, and assessed whether $w$ could be adequately reconstructed.

% Central Limit Theorem
% Confidence interval
% Bound contracts

% law of large numbers --> approach expected value

% \subsection{Poisoning Mitigation}

% SecAgg MITM
% Differential Privacy
% Poisoning Prevention

% Experimental Methods

\chapter{Results}\label{ch:results}
% chktex-file 1
% chktex-file 2
% chktex-file 3
% chktex-file 9
% chktex-file 12
% chktex-file 13
% chktex-file 17
% chktex-file 18
% chktex-file 44
% suppress warning 3 of chktex

% \section{Results}

% noise_tolerance

\begin{figure}%[htp]
  \centering
  \subfloat{%
    \includegraphics[scale=0.43]{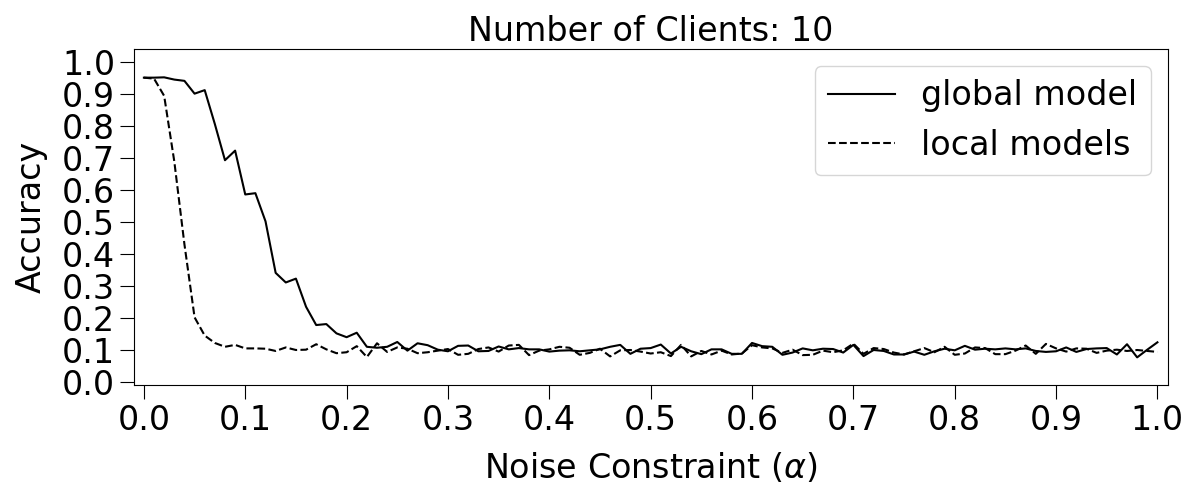}%
  }

  \subfloat{%
  \includegraphics[scale=0.43]{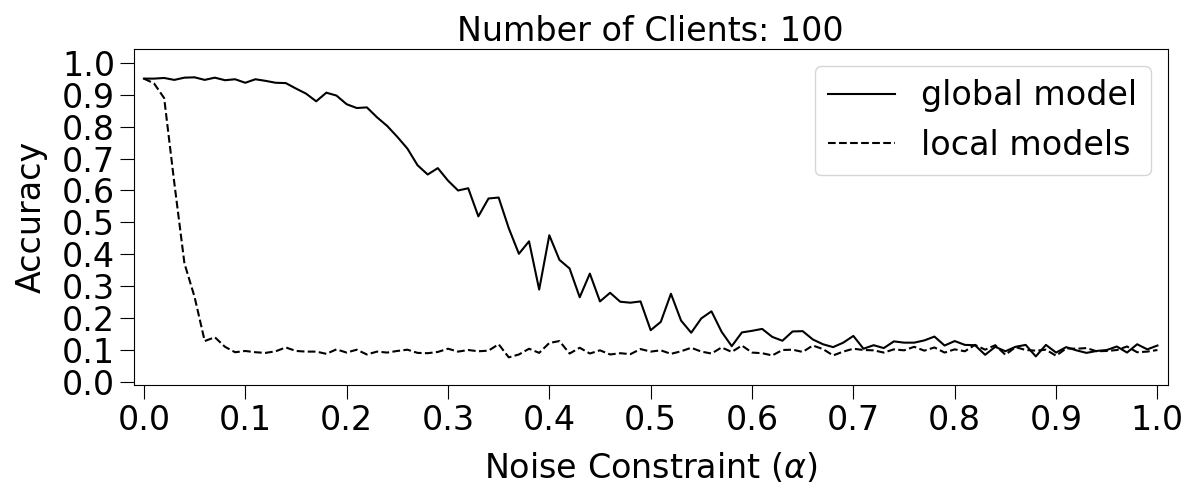}%
  }

  \subfloat{%
  \includegraphics[scale=0.43]{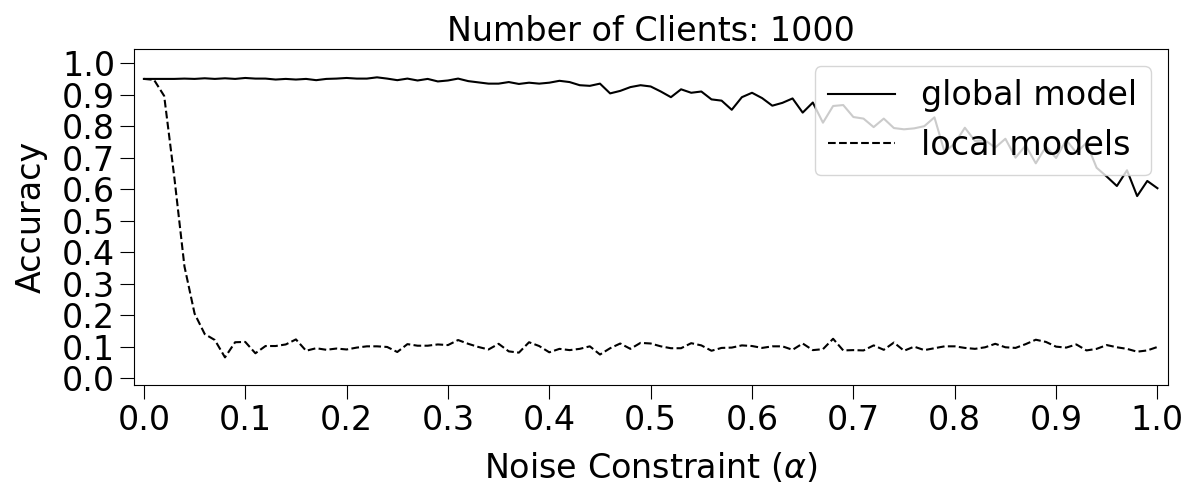}%
  }

  % \caption{Effect of Number of Clients on Noise Tolerance}
  \caption[Effect of Number of Clients on Noise Constraint ($\alpha$) Tolerance with respect to Accuracy]{Effect of Number of Clients on Noise Constraint ($\alpha$) Tolerance with respect to Accuracy. Results were averaged across 10 trials with a pretrained ResNet-20 model on the CIFAR-10 dataset.}\label{fig:alpha-vs-clients}
\end{figure}

Figure~\ref{fig:alpha-vs-clients} shows the effect that changing the amount of noise, indicated by the noise constraint $\alpha$, had on the accuracy of a masked model. The results shown in Figure~\ref{fig:alpha-vs-clients} were averaged across 10 separate trials with a pretrained ResNet-20 model on the CIFAR-10 dataset. In all cases, the weights of the local models were masked with uniform noise within $[-\alpha, \alpha]$ prior to obtaining the accuracy of local models as well as the accuracy of the aggregated global model. It is noteworthy that CIFAR-10 contains 10 classes or types of images, and as such a model guessing at random would be expected to classify the images with an accuracy of $10\%$. This was consistent with the observed results, as local models masked with relatively large values of $\alpha$ still produced a $10\%$ accuracy. This phenomenon is specific to the dataset used.
% TODO: add citation for resnet-20

A significant decrease in accuracy was observed for relatively low values of $\alpha$ for local models. The aggregated global model, however, was able to tolerate higher values of $\alpha$ in all circumstances, with an increase in the number of clients corresponding to an increase in the value of $\alpha$ that could be tolerated for the global model. The rate at which accuracy decayed with increased values of $\alpha$ was also reduced by an increase in the number of clients contributing to the global model, with accuracy diminishing with increased $\alpha$ much more gradually for 1000 clients than compared to 10 clients. It was also observed that for as little as 10 clients, it was possible to obtain values for $\alpha$ that did not compromise accuracy of the global model, but resulted in masked local models having minimal accuracy.

Figure~\ref{fig:alpha-vs-clients} demonstrates that by increasing the number of clients used in aggregation, more noise can be applied to the weights of local models without affecting the accuracy of the aggregated global model. Additionally, it shows that small amounts of noise can completely destroy the accuracy of local models, which presumably preserves the privacy of their training data.

% DLG

\begin{figure}
  \centering
  \subfloat[No Noise Added, $\alpha = 0$]{%
    \includegraphics[scale=0.23]{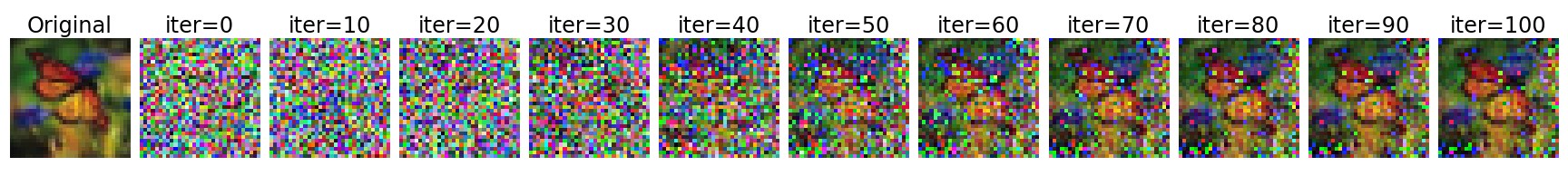}
  }

  \subfloat[$\alpha = 0.01$]{%
    \includegraphics[scale=0.23]{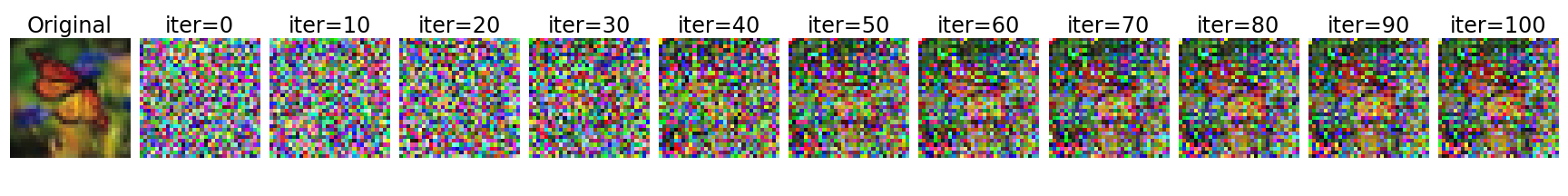}
  }
  
  \subfloat[$\alpha = 0.05$]{%
    \includegraphics[scale=0.23]{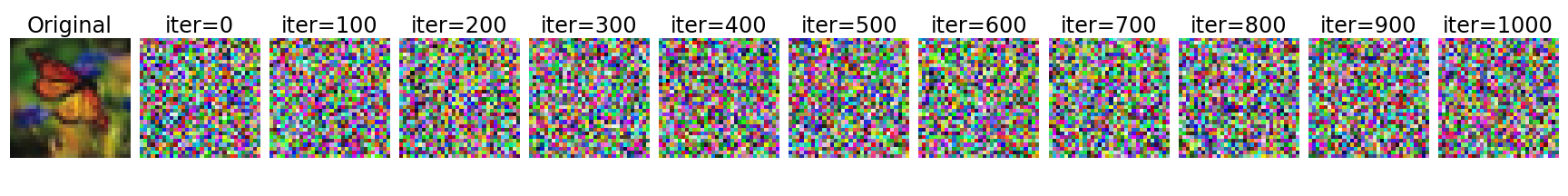}
  }

  \subfloat[$\alpha = 0.1$]{%
    \includegraphics[scale=0.23]{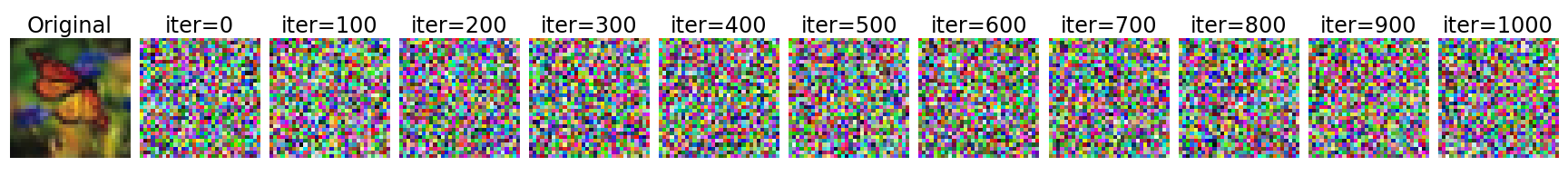}
  }

  \caption[Effect of Noise Constraint ($\alpha$) on DLG Reconstruction Attack Success]{Effect of Noise Constraint ($\alpha$) on DLG Reconstruction Attack Success. The target model was trained for one round on the original image, which was taken from the CIFAR-100 training dataset, prior to the model being masked. The attacker had access to the masked model and gradients, and attempted to reconstruct the original image.}\label{fig:dlg-results}
\end{figure}

Figure~\ref{fig:dlg-results} shows the effect of noise on the capability of performing a DLG Reconstruction Attack, as presented by Zhu et al.\cite{deep-leakage-from-gradients} The original image was an example image taken from the CIFAR-100 training dataset, from which a model was trained for one round, prior to its weights being masked by uniform noise constrained by $\alpha$. The subsequent images were produced by the DLG attack after a given number of iterations.

For $\alpha = 0$, or no noise added, the DLG algorithm was able to reconstruct most of the original example training image after 100 iterations, indicating that the DLG attack was working as expected. However, a relatively small amount of noise used to mask the model, with $\alpha = 0.01$, provided a noticable reduction in the effectiveness of the DLG attack within 100 iterations. Increasing $\alpha$ to $0.05$ completely prevented DLG from successfully reconstructing the original image, even after 1000 iterations of the algorithm. This result remained consistent for higher values of $\alpha$, such as $0.1$.

Figure~\ref{fig:alpha-vs-clients} shows that the accuracy of local models decreased to the minimum expected accuracy by $\alpha = 0.1$, and that accuracy decreased sharply between $\alpha = 0$ and $\alpha = 0.1$. Combined with the results of Figure~\ref{fig:dlg-results}, it can be concluded that the effectiveness of the DLG attack in reconstructing original training data corresponds with the accuracy of the local model. This implies that 100 clients used in aggregation allow for more than enough noise to be added to prevent the server from implementing a DLG attack on local models, without any compromise to accuracy of the global model. Figure~\ref{fig:dlg-results} also implies that the DLG attack is potentially ineffective in the presence of small amounts of noise, implying that differential privacy alone could be an effective defense from this type of attack.

% \begin{figure}
  % \centering
  % optional todo: Insert multi-round dlg here
%
  % \caption{asdf}\label{fig:multiround_dlg}
% \end{figure}

\begin{figure}%[htp]
  \centering
  \subfloat{%
    \includegraphics[scale=0.43]{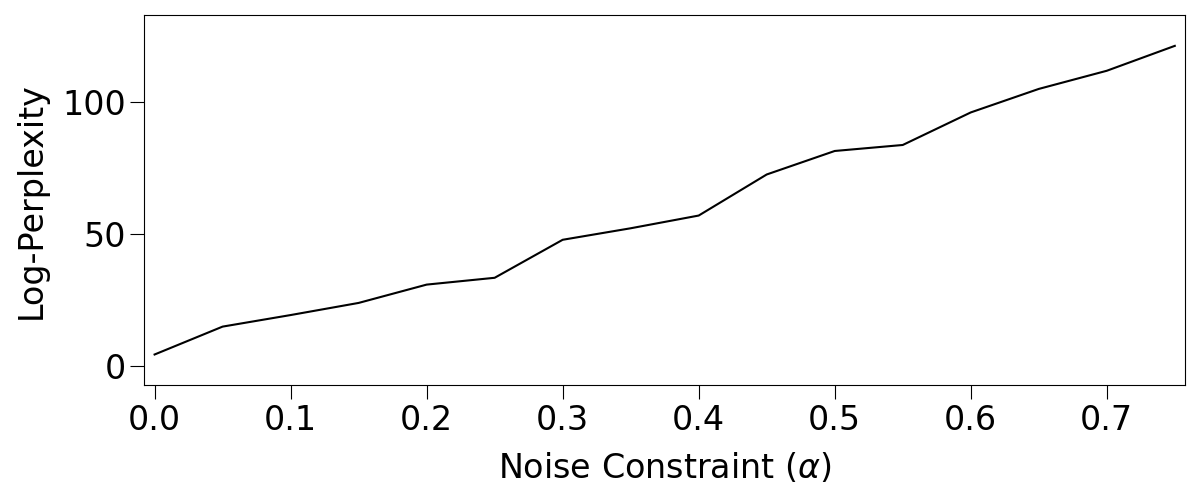}%
  }

  % \caption{Effect of Number of Clients on Noise Tolerance}
  \caption[Effect of Noise Constraint ($\alpha$) on Log-Perplexity of the WikiText Test Dataset for the GPT-2 Language Model]{Effect of Noise Constraint ($\alpha$) on Log-Perplexity of the WikiText Test Dataset for the GPT-2 Language Model. Lower log-perplexity indicates a stronger ability of the model to predict text based on its training data.}\label{fig:alpha-gpt2}
\end{figure}

% perplex
Figure~\ref{fig:alpha-gpt2} demonstrates the effect of increasing the noise constraint $\alpha$ on log-perplexity for a masked language model. The log-perplexity was computed from the negative log-likelihoods of model output versus test data, using a fully trained GPT-2 language model and the WikiText Test Dataset. A lower log-perplexity indicates the model is more capable of predicting test data, thereby making it more vulnerable to the privacy attack proposed by Carlini et al.\cite{secret-sharer} Figure~\ref{fig:alpha-gpt2} shows that log perplexity increases steadily with increased $\alpha$, as expected, implying that the privacy of local language models can be protected with our method, with higher values of $\alpha$ offering more protection. The actual values of log-perplexity are specific to the model and dataset, and thus we recommend using the highest possible $\alpha$ with tolerable global accuracy to err on the side of safety. Additionally, log-perplexities were computed for values of $\alpha$ varied between $0$ and $1$, but became infinitely large for values of $\alpha$ of $0.8$ and above. We assume this was due to the masked model being completely unable to predict text from the test dataset.
% TODO: add citation for GPT-2

\begin{figure}
  % \centering
  % \captionsetup[subfloat]{farskip=2pt, captionskip=1pt}
    % \includegraphics[scale=0.30, valign=t]{gan_orig_data}
  \subfloat[Original Data]{%
    \includegraphics[scale=0.30, valign=t]{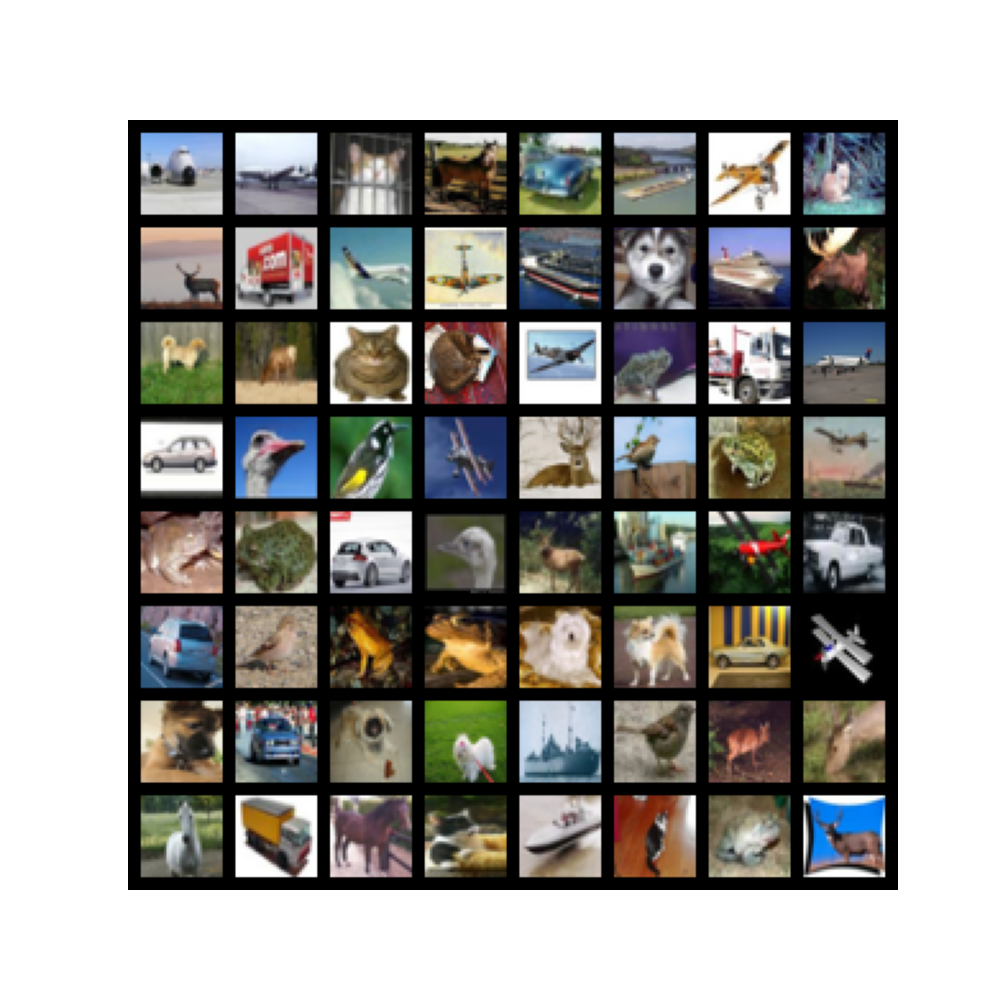}
  } \hspace{-25pt} 
  \subfloat[Regular GAN]{%
    \includegraphics[scale=0.30, valign=t]{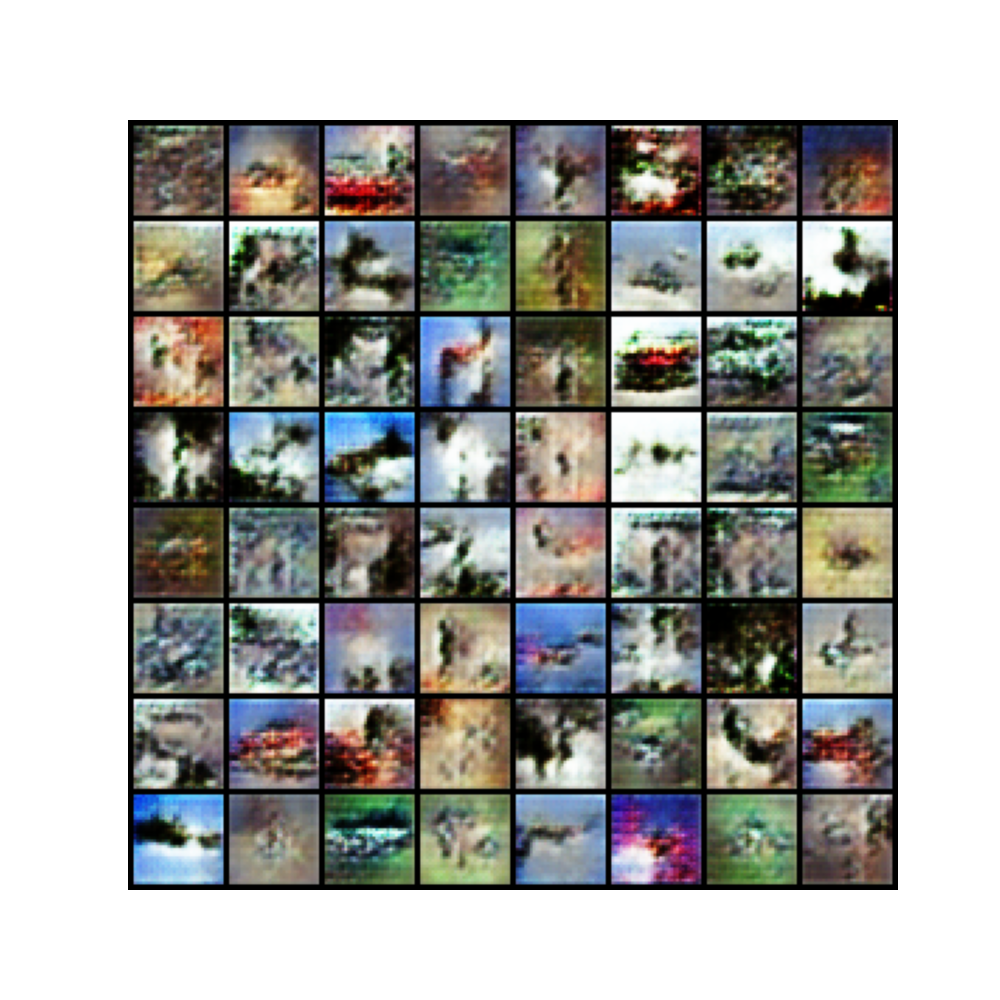}
  }

  \subfloat[Masked Discriminator $\mathcal{D}$ ($\alpha = 0.01$)]{%
    \includegraphics[scale=0.30, valign=t]{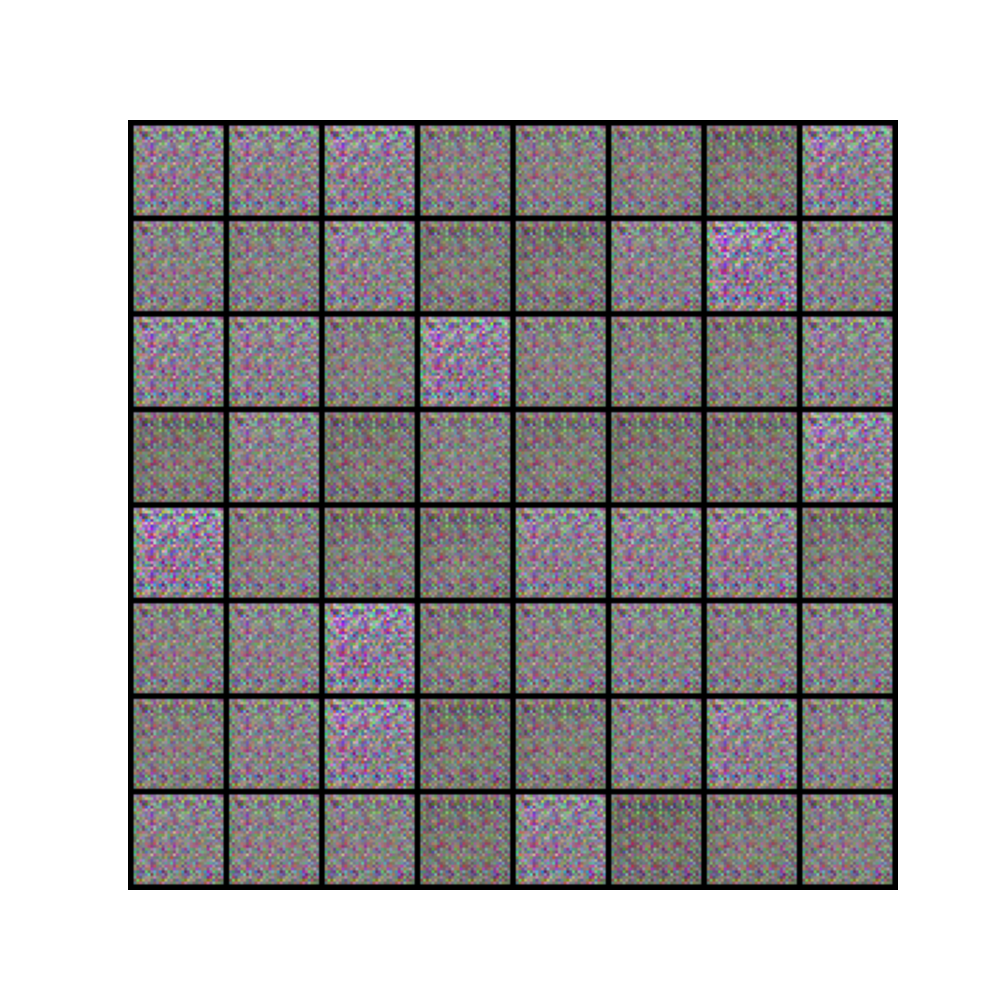}
  } \hspace{-25pt}
  \subfloat[Pretrained Discriminator $\mathcal{D}$ ($t = 5$)]{%
    \includegraphics[scale=0.30, valign=t]{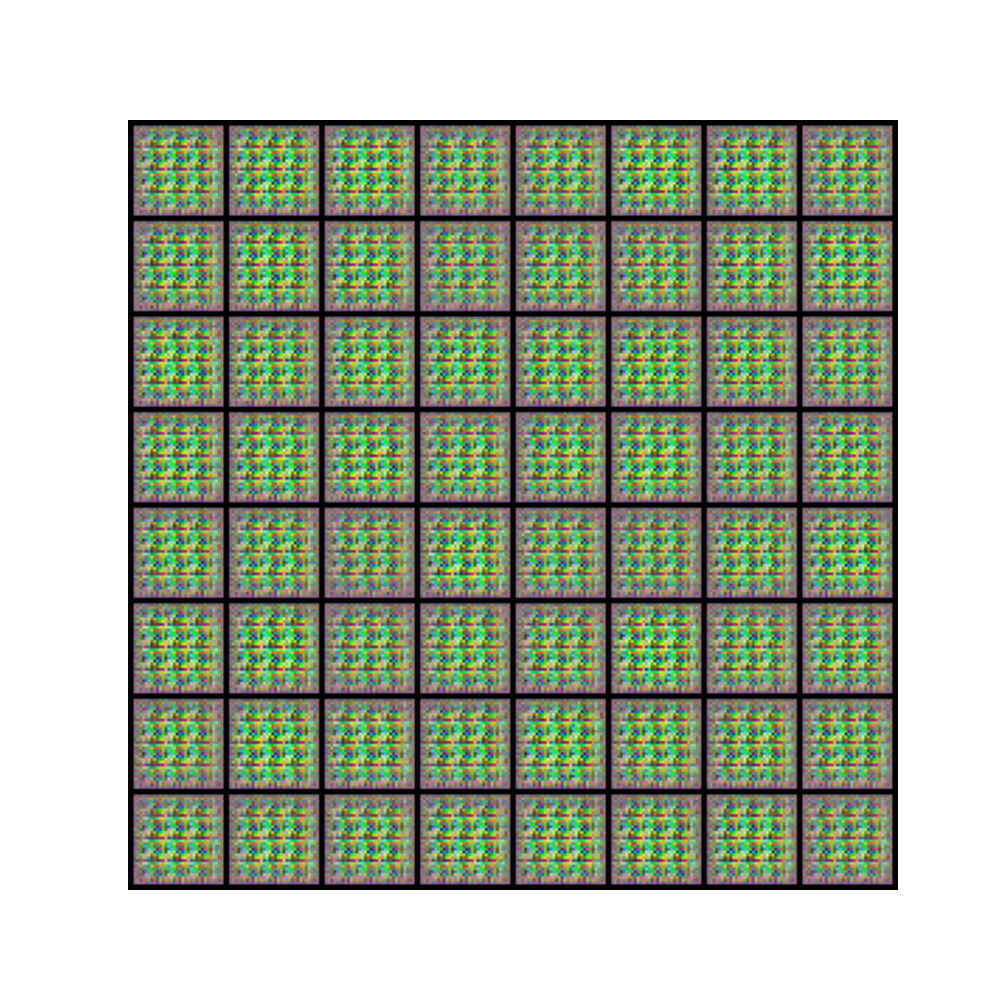}
  }

  \caption[The Instability of the GAN Attack]{The Instability of the GAN Attack. (a) Shows original data from the CIFAR-10 dataset. (b) Shows data generated from a DCGAN on CIFAR-10. (c) Shows data generated after first masking the discriminator $\mathcal{D}$ during training with $\alpha = 0.01$. (d) Shows the effect of first pretraining the GAN discriminator $\mathcal{D}$ for 5 epochs ($t = 5$) prior to training the generator $\mathcal{G}$, after which $\mathcal{G}$ is prevented from influencing the weights of $\mathcal{D}$.}\label{fig:gan_results}
\end{figure}

Figure~\ref{fig:gan_results} demonstrates how the GAN attack by Hitaj et al.\@\cite{deep-models-under-gan} would fail with masked models. Part (b) shows data generated from a DCGAN trained normally on CIFAR-10 data for 5 epochs, which shows some visible similarity to the original training data as shown in Part (a). Part (c) shows the images generated from a generator trained on a discriminator masked with uniform noise with $\alpha = 0.01$, which failed to produce any resemblance to the original data of Part (a). Part (d) also shows a failure to train an effective generator on a discriminator model pretrained for 5 epochs, indicating that the GAN attack would fail on a global model after aggregation or if initiated in later training rounds, as aggregation can occur on multiple local models after several local epochs.

  % the Effects of masking GAN discriminator ($\alpha = 0.01$) or pretraining the discriminator a few rounds on GAN stability.}\label{fig:gan_results}

\FloatBarrier{}

\chapter{Discussion}\label{ch:discussion}
% chktex-file 1
% chktex-file 2
% chktex-file 3
% chktex-file 9
% chktex-file 12
% chktex-file 13
% chktex-file 17
% chktex-file 18
% chktex-file 44
% suppress warning 3 of chktex

% \section{Discussion}

We have shown that secure aggregation is fundamentally insecure in the case of a fully malicious server in the absence of a trusted third party. Consequently, one may conclude that secure aggregation ultimately requires that clients trust the server, which would make federated learning unnecessary, or trust a third party to verify that the server is behaving correctly as claimed. However, a closer inspection of secure aggregation protocols reveals that this creates an enormous burden on any such trusted third party (assuming the trusted third party is not itself malicious or in collusion with the server, which may actually be in the interest of the trusted third party). For instance, at every communication round, it is always the server's job to declare what clients dropped connection. If the trusted third party were to receive the messages sent by the clients instead, then the trusted third party would essentially become the server. If this is not the case, though, then the trusted third party is unable to verify that the server is telling the truth as to which clients dropped connection. By being able to declare which clients dropped connection, the server is able to select which clients ultimately remain in a training round. This gives the server the power to always create a majority of Sybil clients owned by the server, which can make compromising Shamir secret sharing or performing a man-in-the-middle attack possible. This then puts the burden of proof on the trusted third party to verify that clients are not controlled by the server. This challenge bears some similarity to those of certificate authorities in the context of web security, except that, to our knowledge, this task would be conceptually more difficult if not impossible, given that fake Sybil clients could behave just like regular clients. Even in cases where identity verification were to be performed on every client, there would be no conceivable way for the trusted third party to verify that clients were not in collusion with the server. Consequently, not only would the trusted third party have to be trusted, but the majority of clients as well, in a situation where the server can potentially generate and register many fake Sybil clients.
% ... why we can't rely on trusted third parties (certificate authorities) (never mind)

% Could secure aggregation work with decentralized learning?
One scenario that remains largely unexplored with respect to secure aggregation is its application to a decentralized learning context. Decentralized learning differs from federated learning in that clients coordinate communication among themselves and that the training process itself does not involve a centralized server. Rather than a single global model being maintained thoughout the training process, clients reach consensus by converging to a consistent overall model.\cite{advances-and-open-problems-FL} While decentralized learning faces some challenges different from federated learning, an application of secure aggregation to decentralized learning could potentially be secure from a man-in-the-middle Sybil attack, because no individual entity would be able to control client participation or communication. However, it is worth noting that we did not find any compatible secure aggregation algorithm in the literature, as all forms of secure aggregation relied on trust being placed in a central authority. Furthermore, such a method of secure aggregation, were it to be developed, would likely be computationally less efficient from our method, without necessarily being more secure.

% circumstances where secure aggregation could be useful: protecting privacy of previously trained models if the server becomes compromised in the future (provided secure aggregation was implemented correctly)
% However, the same is true for our method, and once the server is compromised and becomes malicious, secure aggregation can no longer be reliable
The only circumstance in which secure aggregation could be preferable to our method would be if a higher amount of noise (e.g., a higher $\alpha$) were needed to mask client models than could be tolerated with our method, and it could be assured that the server was not malicious. In such circumstances, secure aggregation could still be used alongside our method, although to our knowledge no attacks exist on data privacy capable of leaking data from local neural networks masked with our method with reasonably high levels of $\alpha$ (e.g., $\alpha = 0.5$). As such, we conclude that secure aggregation is unnecessary.

% multi-round attacks on secure aggregation are actually not as dangerous as they may seem, because they can only be implemented late in training, at which point the gradient differences between training rounds is smaller. Attacks such as DLG are much more difficult to implement when the gradients are smaller.
So et al.\@ discussed potential privacy concerns that could arise due to similarities between updates once the global model begins to converge, and discussed this in the context of many forms of secure aggregation being insecure as such methods did not take this privacy risk into consideration.\cite{multi-round-secagg} However, our investigation into privacy attacks such as DLG revealed that DLG was more effective at leaking data when gradient updates were larger, (i.e., at the beginning of training). Once the global model starts to converge, it changes less between updates, meaning that individual updates result in smaller changes in gradient, thereby making the DLG attack less stable and effective in later stages of training. Consequently, we are skeptical that model similarity between updates is as strong of a privacy concern as So et al.\@ implied.

% DLG affected by even a small amount of noise; current privacy attacks are brittle (for images)
Nevertheless, based on our experimental results, the DLG attack was only effective in the presence of little or no noise with our method ($\alpha < 0.05$), which, based on Figure~\ref{fig:dlg-results}, implies that as little as $10$ clients per aggregation step could be sufficient to prevent the DLG attack from being a viable concern. In practice, many federated learning systems perform aggregation on the order of hundreds or thousands of clients.\cite{advances-and-open-problems-FL} This implies that, with our method, the DLG attack is no longer a legitimate privacy concern for federated learning scenarios.

% this does not mean future attacks will be as brittle
It is possible, however, for there to be future development in the area of privacy attacks on neural networks. Under such circumstances, it is conceivable that higher values of $\alpha$ will be needed to protect the privacy of clients than is currently the case. For this reason, we recommend using the highest tolerable $\alpha$ with our method, with potential tradeoffs between accuracy or the number of clients being taken into consideration. We also recommend semi-honest servers to remove previous information on training updates once training is completed, as this could reduce the risks associated with the server later becoming compromised or malicious at a later date, at which time stronger attacks on data privacy may be available. 
% Note that the effect of increasing $\alpha$ on accuracy can be reduced by increasing the number of clients, which would conceivably become increasingly feasible with future improvements to networking.
% TODO: number of clients can also increase with network improvements
% if you increase the number of clients, you don't need to worry about a decrease in accuracy, which can be done with increased network improvements

% Why the GAN attack by Hitaj et al. is impractical for federated learning
% attacker doesn't train every round; there are many other clients
Figure~\ref{fig:gan_results} indicates that pretraining the discriminator of a GAN for several rounds can prevent the generator from ever converging. This implies that for a successful GAN attack as demonstrated by Hitaj et al.\@\cite{deep-models-under-gan} to be achieved, the attacker would have to initiate the attack within the very first round of training, and continue the attack throughout the entire training process. Furthermore, the attacker would have to train the generator at a relatively similar pace to the discriminator, implying that there would have to be very few clients contributing to the global model in addition to the attacker. These assumptions do not typically hold in a federated learning setting. Thus, it would be very easy for the server to prevent malicious clients from performing the GAN attack by first pretraining the global model on publicly available data to prevent convergence, in addition to ensuring multiple clients contribute to the global model each training round, and selecting different clients each round. In the case of a malicious server, our defense ensures that local models cannot be viewed by the server prior to aggregation with a sufficiently high $\alpha$ and a sufficient number of clients, as shown in Figure~\ref{fig:gan_results}, where masking the discriminator with $\alpha = 0.01$ prevented the GAN from training. Furthermore, by expanding the number of clients that contribute to the global model per round, the global model is effectively trained on more data prior to being accessible to a generator model, making the GAN attack even more impractical given that it failed to train with a pretrained discriminator. Our results, as exemplified by Figure~\ref{fig:gan_results}, indicate that the application of our method causes the GAN attack by Hitaj et al.\@ to not be a viable concern for federated learning.  

% Why did the GAN fail
We attribute the failure of the GAN to train on a global model to the fact that if the discriminator is significantly stronger than the generator, then everything the generator initially generates will be classified as fake or not part of the target class. Consequently, it follows logically that the generator does not obtain enough feedback to converge when the discriminator is significantly stronger than the generator, which would be the case when applying the GAN attack of Hitaj et al.\@ to federated learning.

% Carlini et al.'s attack on language models
Figure~\ref{fig:alpha-gpt2} shows a near linear correlation between $\alpha$ and log-perplexity of local language models, with log-perplexity becoming incalculable for $\alpha \geq 0.8$ for a pretrained GPT-2 model on the WikiText test dataset. Given that the Secret Sharer privacy attack by Carlini et al.\@\cite{secret-sharer} requires the calculation and minimization of log-perplexity, it follows that increasing values of $\alpha$ make this attack increasingly difficult on local models, with high values of alpha (e.g., $\alpha \geq 0.8$) making the attack completely impossible. It may be acceptable to tolerate lower levels of $\alpha$, but this will depend on the model being attacked, and on the repetition of sensitive data within the training dataset. Nevertheless, our results indicate that our method can be a viable defense against the Secret Sharer attack on local client models. Furthermore, this is consistent with the results of Carlini et al.\@, who found that differential privacy could be effective at preventing their attack.\cite{secret-sharer}

% Can Carlini's attack work on global language models? According to their research, not if sufficient differentl privacy is used during each training step
Carlini et al.\@ also indicated in their paper that differentially private SGD was effective at mitigating data memorization for language models.\cite{secret-sharer} Given that the Secret Sharer attack is possible on language model after it is fully trained, (e.g., a fully aggregated global model), it follows that differentially private SGD should be employed during the training of language models in addition to our method. Note that our method only aims to protect the privacy of local client models, rather than the global model, which is also the goal of secure aggregation.

% Can our method be used in Async operations?
Although we did not explore the use of our algorithm with asynchronous optimization methods such as FedAsync or FedBuff, our method could still be theoretically applicable to some asynchronous scenarios.\cite{fedasync, fedbuff} With FedAsync, the staleness and mixing parameters reduce the contribution effect of clients that are slower to update the global model. While this would simultaneously reduce the effect of noise for those clients on aggregation, this would not impact the tendency of the noise to average to zero by the central limit theorem. Provided that there were enough clients with non-stale updates contributing to the global model, this should not have a significant effect on the aggregation of the overall global model. Our method could also be applied to FedBuff, provided that enough clients exist within a buffer to achieve acceptable levels of $\alpha$.

% Can our method work with byzantine-tolerant aggregation methods?
Our method, however, would potentially not work as well with some Byzantine-tolerant aggregation methods, such as the Geometric Median or Coordinate-wise Median, as the central limit theorem would not necessarily apply to median-based approaches.\cite{robust-fedavg, trimmed-mean-median, central-limit} However, as Karimireddy et al.\@ demonstrated, median-based Byzantine aggregation methods are vulnerable to a timing attack.\cite{byzantine-history} Methods considered to be more robust, such as Centered Clip, involve the summation of vectors which could contain uniform noise from our method, the effect of which should converge to zero by the central limit theorem.\cite{byzantine-history, central-limit} Furthemore, methods such as Krum, Bulyan, and Coordinate-wise Trimmed Mean aim to achieve Byzantine robustness by excluding outliers, but still make use of averaging in their methods.\cite{krum, bulyan, trimmed-mean-median} Thus, our method should work for such methods, although it may be necessary to further increase the number of clients to account for Byzantine fault tolerance.
% Can poisoning attack defenses work with our method?
Additional defenses for poisoning attacks, such as the use of ensembles or weight pruning, could be performed on the global model after training, and consequently they could be used in addition to our method. Therefore, the use of our method should not hinder the use of defenses against poisoning attacks.

% OPTIONAL: Moved future work into conclusion?

% What are potential bottlenecks for using more clients
% Future work
One potential drawback of our method is it could require more clients to participate within an aggregation round. While this is desirable for privacy purposes, it could ultimately result in slower training times and cause complications in the event of a large number of clients dropping connection. However, our results indicate that current attacks on data privacy are also brittle in the presence of noise, and as a result we believe our method to be applicable to the majority of use cases with federated learning. 
% One potential drawback of our method is it could require more clients to participate within an aggregation round. While this is desirable for privacy purposes, it could ultimately result in slower training times and cause complications in the event of a large number of clients dropping connection. However, our results indicate that current attacks on data privacy are also brittle in the presence of noise, and as a result we believe our method to be applicable to the majority of use cases with federated learning. Future work could involve investigation of our method applied to asynchronous optimization methods, as this would reduce the effect of client dropout, and Byzantine fault tolerant methods, which could reduce the influence of potentially malicious clients trying to manipulate the outcome of the training process via poisoning attacks.

Overall, we believe our method to be a far better alternative to secure aggregation, as noise is added only locally with our method, and there is no risk of a Sybil man-in-the-middle attack or a compromise to secret sharing.

\chapter{Conclusion and Future Work}\label{ch:conclusion}
% chktex-file 1
% chktex-file 2
% chktex-file 3
% chktex-file 9
% chktex-file 12
% chktex-file 13
% chktex-file 17
% chktex-file 18
% chktex-file 44
% suppress warning 3 of chktex

% Future work
Future work could involve investigation of our method applied to asynchronous optimization methods, as this would reduce the effect of slow clients or dropped connections, and Byzantine fault tolerant methods, which could reduce the influence of potentially malicious clients trying to manipulate the outcome of the training process via poisoning attacks. Additionally, the combination of our method with other techniques such as quantization and model pruning could be explored, as that could further reduce communication overhead and potentially reduce the risk of poisoning and reconstruction attacks.

% \section{Conclusion}

In federated learning, secure aggregation is fundamentally insecure in the presence of a fully malicious server, as the server has control over orchestrating the federated learning process and can conceivably perform a Sybil man-in-the-middle attack to reconstruct masks used in secure aggregation, or control a majority of fake clients or Sybils to compromise secret sharing protocols. This can be done without any reliable way for clients to verify that the server has not circumvented secure aggregation. We demonstrate that secure aggregation is irrelevant when the number of clients used in aggregation is increased, as more clients allow for more noise to be added locally without compromising global model accuracy. Additionally, our experimental results indicate that current attacks on data privacy for neural networks cannot withstand even small amounts of noise applied to local client weights. Overall, we conclude that relying on secure aggregation to preserve data privacy both provides a false sense of security and is unnecessary. Furthermore, our alternative of simply scaling up the number of clients along with locally-added noise is far simpler, and cannot be compromised via a compromise to secret sharing or via a man-in-the-middle attack.

% \chapter{Appendix}\label{ch:appendix}
% chktex-file 1
% chktex-file 2
% chktex-file 3
% chktex-file 9
% chktex-file 12
% chktex-file 13
% chktex-file 17
% chktex-file 18
% chktex-file 44
% suppress warning 3 of chktex

\begin{appendices}\label{ch:appendix}
% \appendix\label{ch:appendix}

% \section{Notation}
\chapter{Notation}

\begin{longtable}{r l}\label{tbl:notation}
  $\alpha$ & noise constraint (our method)  \\
  $\xi$ & noise scaling parameter (DP-SGD) \\
  $a$ & a polynomial coefficient or model activation \\
  % $U$ & set of clients  \\
  $w$ & model weights \\
  $b$ & model biases \\
  $\theta$ & model activation functions \\
  $f_w(x)$ & a neural network model, classifier function \\
  $\nabla f_w$ & gradient \\
  $u$ & users or clients \\
  $X,Y$ & IID variables  \\
  $B$ & fraction of Byzantine machines \\
  $x$ & input data  \\
  $y$ & output or data label  \\
  $ \mathbb{D}$ & gradient difference (DLG) \\
  $\nabla_x \mathbb{D}, \nabla_y \mathbb{D}$ & change in gradient difference for $x,y$ \\
  $\zeta, \gamma$ & threshold parameters (MIA, DP-SGD) \\
  $D$ & a dataset  \\
  $\mathcal{L}$ & loss function for computing gradients  \\
  $\mathcal{A}$ & an algorithm or aggregation rule  \\
  $\mathcal{N}$ & normal distribution  \\
  $\mathcal{U}$ & uniform distribution  \\
  $Z$ & standardized normal distribution \\
  $\mu$ & mean  \\
  $\sigma^2$ & variance  \\
  $k$ & threshold for secret sharing \\
  $\bar{x}$ & sample mean \\
  $\rho$ & a 2D polynomial \\
  $m$ & a secret message \\
  $c$ & ciphertext (encrypted message) \\
  $M$ & mask (SecAgg, our method) \\
  $C$ & cost function (MIA) \\
  $\epsilon, \delta$ & arbitrarily small values, DP parameters \\
  $\eta$ & learning rate or step size \\
  $\beta$ & momentum  \\
  $enc$ & encryption function \\
  $dec$ & decryption function \\
  $p, q$ & prime numbers \\
  $e,d$ & encryption / decryption RSA integers \\
  $g$ & Diffie-Hellman generator \\
  $\lambda$ & function for computing RSA keys \\
  % $N,K,L$ & sizes for sets or Big-O Notation \\
  $n$ & set size \\
  % $K$ & random sample size \\
  $h$ & number of sampled clients \\
  $r$ & dimensions \\
  $S$ & a set \\
  $t$ & time step or epoch \\
  $T$ & total number of time steps \\
  $\nu$ & a vector \\
  $s$ & a shared secret \\
  $sg$ & a signature \\
  $pk$ & public key \\
  $sk$ & secret key \\
  $\mathcal{G}$ & GAN generator model \\
  $\mathcal{D}$ & GAN discriminator model \\
  $I$ & group identity \\
  $\tau$ & radius \\
  $\ell$ & length of model updates \\
  & \\
  MITM & man-in-the-middle \\
  IID & independent and identically distributed \\
  LCM & least common multiple \\
  $O$ & Big-O Notation (algorithmic complexity) \\
  SecAgg & Secure Aggregation \\
  DP & Differential Privacy \\
  MPC & Multi-Party Computation \\
  GAN & Generative Adversarial Network \\
  2PC & Two-Party Computation \\
  DCGAN & Deep Convolutional GAN \\
  MIA & Model Inversion Attack \\
  LP & Log-Perplexity \\
  DLG & Deep Leakage from Gradients \\
  SGD & Stochastic Gradient Descent \\
  FedAvg & Federated Averaging \\
  $\Pr$ & Probability \\
  DP-SGD & Differentially Private SGD \\
  $\textsc{Kr}$ & Krum aggregation \\
  RFA & Robust Federated Averaging \\
  TM & Trimmed-Mean Aggregation \\
  PATE & Private Aggregation of Teacher Ensembles \\
  FGSM & Fast Gradient Sign Method \\
  PGD & Projected Gradient Descent \\
  BPDA & Backward-Pass Differential Approximation \\
  LWE & Learning with Errors \\
  EFF & Electronic Frontier Foundation \\
  FLoC & Federated Learning of Cohorts \\
  AI & Artificial Intelligence \\
  GDPR & General Data Protection Regulation \\
\captionlistentry{Notation}
\end{longtable}

% \section{Byzantine-Tolerant Aggregation Techniques}\label{byzantine_techniques}
\chapter{Byzantine-Tolerant Aggregation Techniques}\label{byzantine_techniques}
% krum and stuff

\textbf{Krum} Krum is an aggregation method proposed by Blanchard et al.\@ that aims to approximate averaging but with a provable tolerance to failure among workers. Additionally, they experimentally demonstrated cases where Krum was unaffected by 45\% Gaussian Byzantine workers. Krum aims to approximate the mean after excluding outliers. After excluding $\delta n + 2$ furthest away points for a selected $\delta$, Krum returns the point $x_i$ closest to the mean as follows. Suppose $\mathcal{S} \subset [n]$ of size at least $(n-\delta n - 2)$. Then,

\[ \textsc{Kr}(x_1, \ldots, x_n) = \mathop{\text{argmin}}_{x_i} \mathop{\text{min}}_{\mathcal{S}} \sum_{j\in \mathcal{S}} \Vert x_i - x_j \Vert^2_2 \]

Note that Krum operates with a time complexity of $O(n^2 r)$, where $r$ is the number of dimensions of $n$ data points.\cite{krum}

\textbf{Robust Federated Averaging (RFA) / Geometric Median} Pillutla et al.\@ proposed the use of an approximate geometric median in place of the weighted arithmetic mean. The robust federated averaging or geometric median is defined formally is as follows:

\[ \text{RFA}(x_1, \ldots, x_n) = \mathop{\text{argmin}}_\upsilon \sum_{i=1}^{n} \Vert \upsilon - x_i \Vert_2 \]

They demonstrated that RFA is robust to data corruption as well as a poisoning attack in which the weights are set to push the weighted arithmetic mean towards the negative of what it would have been, which aimed to obstruct convergence. RFA achieves this robustness at three times the communication overhead compared to the regular arthimetic mean.\cite{robust-fedavg}

\textbf{Bulyan} Mhamdi et al.\@ showed that while Krum and RFA were shown to converge despite Byzantine failures, it was possible for an attacker to force them to converge to an ineffectual model due to the existence of many possible local minima. They proposed Bulyan as an alternative aggregation technique that would be robust to such an attack. Bulyan works by ensuring that each coordinate is agreed on by a majority of vectors after performing a Byzantine-resilient aggregation rule $\mathcal{A}$ (i.e.\@ Krum or RFA). For any Byzantine aggregation rule $\mathcal{A}$, $\mathcal{A}$ is performed to select coordinates among proposed vectors, after which those coordinates are moved to a selection set $\mathcal{S}$ and removed from the original set. This step is performed recursively while $\vert \mathcal{S} \vert < \gamma$ for a predetermined $\gamma = n-2d \geq 2d + 3$, where $d \in \left[0, \frac{\pi}{2}\right] \times \{0, \ldots, n \}$. This ensures $\mathcal{S} = (S_1, \ldots, S_\gamma)$ contains a majority of non-Byzantine gradients, which can be aggregated to compute the final gradient $\nabla f_w$. The coordinates of the resulting gradient $\nabla f_w$ can be computed as follows:

\[ \forall i \in [1 \ldots \ell], \nabla f_w[i] = \frac{1}{\zeta} \sum_{\nu_1 \in \mathcal{M}[i]} \nu_1[i] \]

where $\zeta = \gamma - 2d \geq 3$ and: 

\[ \mathcal{M}[i] = \mathop{\text{argmin}}_{R \subset \mathcal{S}, \vert R \vert = \zeta} \left(\sum_{\nu_1\in \mathcal{R}} \vert \nu_1[i] - \text{median}[i] \vert \right) \]

\[ \text{median}[i] = \mathop{\text{argmin}}_{b=Y[i], Y\in \mathcal{S}} \left(\sum_{\nu_2\in \mathcal{S}} \vert \nu_2[i] - b \vert \right) \]

Bulyan was shown to have a computational complexity of $O(n^2 \ell)$ if $\mathcal{A}$ were to be Krum or RFA.\cite{bulyan}

\textbf{Coordinate-wise Trimmed Mean} Yin et al.\@ proposed the trimmed mean, which first involves removing the largest and smallest $\zeta$ fraction of elements prior to taking the mean for $\zeta \in [0, \frac{1}{2})$. Formally, it is defined as follows, where $\Pi_j$ denotes the sorting of each coordinate value $j$, and the $\zeta n$ largest and smallest values are excluded:

\[ TM(x_1, \ldots, x_n) = \dfrac{1}{n - 2\zeta n} \sum_{i=\zeta n}^{n-\zeta n} \left[x_{\Pi_j(i)} \right]_j \]

The trimmed mean was shown to achieve an order-optimal error rate of $\mathcal{\widetilde{O}}\left(\frac{B}{\sqrt{n}} + \frac{1}{\sqrt{nh}}\right) $, where $B$ is the fraction of machines that are Byzantine for $h$ machines, although for strongly convex quadratic problems, it achieves an error rate of $\mathcal{\widetilde{O}} \left(\frac{B}{\sqrt{n}} + \frac{1}{\sqrt{nh}} + \frac{1}{n} \right)$.\cite{trimmed-mean-median}

\textbf{Coordinate-wise Median} Yin et al.\@ also proposed and assessed the use of a coordinate-wise median as an aggregation technique for federated learning. The coordinate-wise median is essentially a median performed on each coordinate of a vector, formally defined as $\nu = med\{ x^i : i \in [n] \}$ for a vector with the j-th coordinate being $\nu_j = med\{x_j^i : i \in [n] \}$, for vectors $x^i \in \mathbb{R}^r, i \in [n]$ for each $j \in [r]$ for $r$-dimensions. Essentially, it is the one-dimensional median performed along each element of a matrix. The coordinate-wise median could achieve an error rate of $\mathcal{\widetilde{O}} \left(\frac{B}{\sqrt{n}} + \frac{1}{\sqrt{nh}} + \frac{1}{n} \right)$ for $h$ machines, similar to that of the coordinate-wise trimmed mean.\cite{trimmed-mean-median}

\textbf{Adaptive FedAvg} Mu$\tilde{\text{n}}$oz-Gonzalez et al.\@ used a Hidden Markov model to assess the quality of model updates and block potentially bad updates, while being more computationally efficient than previous robust aggregation techniques such as Krum and coordinate-wise median. However, their method requires computation of the similarity between each client's local weights and the aggregated global weight, which thereby requires that the server can access or approximate local client weights. Therefore this method is not compatible with secure aggregation, although it may work with differential privacy to some degree.\cite{adaptive-fedavg}

\textbf{Centered Clip} Karimireddy et al.\@ demonstrated that several previously proposed Byzantine-tolerant aggregation techniques that are based on the median are vulnerable to timed attacks across rounds, in which perturbations are kept within the variance of good gradients, as well as being vulnerable to unbalanced data and data bias. For instance, if there are $n$ random variables that are $\pm 1$ for an odd $n$, the mean will be 0 but methods such as Krum, coordinate-wise median, and Bulyan will return either of $\pm 1$. Additionally, if data is unbalanced and a majority of data corresponds to a particular class, the median of the gradients will likely correspond to gradients of that class, thereby ignoring all other classes in optimization. To counteract these issues, they proposed two techniques, one being an iterative clipping procedure and the other involving worker momentum. The centered-clipping procedure can be defined as follows, starting from some initial vector $\nu_0$ for iterations $\ell \geq 0$ and radius $\tau$:

\[ \nu_{\ell+1} = \nu_{\ell} + \frac{1}{n} \sum_{i=1}^{n}(x_i - \nu_\ell) \text{min}\left(1, \frac{\tau_\ell}{\Vert x_i - \nu_\ell \Vert} \right) \]

Their second technique involved the use of worker sharing momentum with the server in place of gradients, as non-timing attack perturbations would not effectively reduce variance between rounds, making Byzantine perturbations easier to detect. The worker would update momentum like so:

\[ \beta_{t,i} = (1 - \zeta_t) \nabla f_{w,i}(x_{t-1}) + \zeta_t \beta_{t-1,i}\]

The server would aggregate the momenta and update the weights as follows, where $\textsc{agg}$ is a Byzantine-resilient aggregation rule such as centered-clipping:

\[ \beta_{t} = \textsc{agg}(\beta_{t,1}, \ldots, \beta_{t,n})\]
\[ x_t = x_{t-1} - \eta_t \beta_t\]

Based on their results, their method outperformed previous methods, and was more computationally efficient as $O(n)$. Additionally, their method is compatible with secure aggregation and asynchronous optimization methods.\cite{byzantine-history} % use this citation to write explanations

\chapter{Model Code}\label{model_arch}

Below is the code for the neural network architectures used in each experiment. Experiments were run using Python 3.8 on a NVIDIA GeForce RTX 2080 Ti GPU.\@ The rest of the code will later be made available at \url{https://github.com/johngilbert2000}

\section{Noise Tolerance (ResNet-20)}

The pretrained ResNet-20 models were obtained from tf2cv: \url{https://github.com/osmr/imgclsmob}

\begin{lstlisting}[language=Python]
import tensorflow as tf
from tf2cv.model_provider import get_model
from tensorflow.keras.datasets import cifar10, cifar100

def generate_model():
  return get_model("resnet20_cifar10", pretrained=True)
\end{lstlisting}

\section{DLG Attack (LeNet)}

A LeNet architecture was used for the DLG attack.

\begin{lstlisting}[language=Python]
import torch
import torch.nn as nn

class LeNet(nn.Module):
  def __init__(self):
    super(LeNet, self).__init__()
    self.body = nn.Sequential(
      nn.Conv2d(3, 12, kernel_size=5, padding=5//2, stride=2),
      nn.Sigmoid(),
      nn.Conv2d(12, 12, kernel_size=5, padding=5//2, stride=2),
      nn.Sigmoid(),
      nn.Conv2d(12, 12, kernel_size=5, padding=5//2, stride=1),
      nn.Sigmoid(),
      nn.Conv2d(12, 12, kernel_size=5, padding=5//2, stride=1),
      nn.Sigmoid(),
    )
    self.fc = nn.Sequential(
      nn.Linear(768, 100)
    )
    
  def forward(self, x):
    out = self.body(x)
    out = out.view(out.size(0), -1)
    out = self.fc(out)
    return out
\end{lstlisting}

\section{Log-Perplexity (GPT-2)}

A pretrained GPT-2 model was obtained from the Hugging Face library: \url{https://huggingface.co/}

\begin{lstlisting}[language=Python]
from transformers import GPT2LMHeadModel, GPT2TokenizerFast
from datasets import load_dataset
import torch

def get_model_and_tokenizer(model_id="distilgpt2"):
  model = GPT2LMHeadModel.from_pretrained(model_id).to(DEVICE)
  tokenizer = GPT2TokenizerFast.from_pretrained(model_id)
  return model, tokenizer

def load_data():
  data = load_dataset("wikitext", "wikitext-2-raw-v1")
  return data["train"]["text"], data["validation"]["text"], data["test"]["text"]
\end{lstlisting}

\section{GAN Attack (DCGAN)}

The DCGAN architecture used for the GAN experiments is shown below.

\begin{lstlisting}[language=Python]
import torch
import torch.nn as nn

class Generator(nn.Module):
  def __init__(self, input_size=100, 
              feature_size = 64, num_channels=3):
    super(Generator, self).__init__()
    self.input_size = input_size
    self.main = nn.Sequential(
      nn.ConvTranspose2d(self.input_size, feature_size * 8,
        4, 1, 0, bias=False),
      nn.BatchNorm2d(feature_size * 8),
      nn.ReLU(True),
      # state size. (feature_size*8) x 4 x 4
      nn.ConvTranspose2d(feature_size * 8, feature_size * 4,
        4, 2, 1, bias=False),
      nn.BatchNorm2d(feature_size * 4),
      nn.ReLU(True),
      # state size. (feature_size*4) x 8 x 8
      nn.ConvTranspose2d( feature_size * 4, feature_size * 2,
        4, 2, 1, bias=False),
      nn.BatchNorm2d(feature_size * 2),
      nn.ReLU(True),
      # state size. (feature_size*2) x 16 x 16
      nn.ConvTranspose2d( feature_size * 2, feature_size,
        4, 2, 1, bias=False),
      nn.BatchNorm2d(feature_size),
      nn.ReLU(True),
      # state size. (feature_size) x 32 x 32
      nn.Conv2d( feature_size, num_channels, 3, 
        1, 1, bias=False), # use with 32x32 imgs
      nn.Tanh()
    )

  def forward(self, input):
    return self.main(input)
  
class Discriminator(nn.Module):
  def __init__(self, feature_size=64, num_channels=3):
    super(Discriminator, self).__init__()
    self.main = nn.Sequential(
      # input is (num_channels) x 64 x 64
      nn.Conv2d(num_channels, feature_size, 
        4, 2, 1, bias=False),
      nn.LeakyReLU(0.2, inplace=True),
      # state size. (feature_size) x 32 x 32
      nn.Conv2d(feature_size, feature_size * 2,
        4, 2, 1, bias=False),
      nn.BatchNorm2d(feature_size * 2),
      nn.LeakyReLU(0.2, inplace=True),
      # state size. (feature_size*2) x 16 x 16
      nn.Conv2d(feature_size * 2, feature_size * 4,
        4, 2, 1, bias=False),
      nn.BatchNorm2d(feature_size * 4),
      nn.LeakyReLU(0.2, inplace=True),
      # state size. (feature_size*4) x 8 x 8
      nn.Conv2d(feature_size * 4, feature_size * 8, 
        4, 2, 1, bias=False),
      nn.BatchNorm2d(feature_size * 8),
      nn.LeakyReLU(0.2, inplace=True),
      nn.Conv2d(feature_size * 8, 1,
        2, 1, 0, bias=False),
      nn.Sigmoid()
    )
\end{lstlisting}

\end{appendices}

%------------------------------------
% Thesis Body -- end
%------------------------------------

% \bibliographystyle{IEEEtran}
% \bibliography{references}

\printbibliography
\end{document}